%% file: TSE22-privacy-main-majRev-wh.tex
\documentclass[10pt,journal,compsoc]{IEEEtran}
\usepackage[utf8]{inputenc}
\usepackage{booktabs} 
\usepackage{lscape}
\usepackage{hyperref}
\usepackage{color}
\usepackage{soul}
\usepackage[nocompress]{cite}
\usepackage{booktabs}
\usepackage{longtable}
\usepackage{graphicx}
\usepackage{multirow}
\usepackage{multicol}
\usepackage{mathrsfs}
\usepackage{amssymb}
\usepackage{amsbsy}
\usepackage{amsmath}
\usepackage{tcolorbox}
\usepackage{balance}
\usepackage{microtype}
\usepackage{array}
\usepackage{longtable}
\usepackage{indentfirst}
\usepackage{pgfplots}
\usepackage{enumitem}
\usepackage{algorithmic}
\usepackage{textcomp}
\usepackage{xcolor}
\usepackage{eurosym} 
\usepackage[numbers]{natbib}

\begin{document}
%
\title{Mining and Classifying Privacy and Data Protection Requirements in Issue Reports}

\author{Pattaraporn Sangaroonsilp, Hoa Khanh Dam, Morakot Choetkiertikul, Chaiyong Ragkhitwetsagul, \\Aditya Ghose
\IEEEcompsocitemizethanks{
	\IEEEcompsocthanksitem P. Sangaroonsilp, H. K. Dam and A. Ghose are with the School of Computing and Information Technology, Faculty of Engineering and Information Sciences, University of Wollongong, Australia. \hfil\break Email: ps642@uowmail.edu.au and \{hoa, aditya\}@uow.edu.au
	\IEEEcompsocthanksitem M. Choetkiertikul and C. Ragkhitwetsagul are with the Faculty of Information and Communication Technology, Mahidol University, Thailand. \hfil\break Email: \{morakot.cho, chaiyong.rag\}@mahidol.ac.th}}

%
%

\markboth{Journal of \LaTeX\ Class Files,~Vol.~14, No.~8, August~2015}%
{Shell \MakeLowercase{\textit{et al.}}: Bare Demo of IEEEtran.cls for Computer Society Journals}
%



\IEEEtitleabstractindextext{%

\begin{abstract}
	
Digital and physical footprints are a trail of user activities collected over the use of software applications and systems. As software becomes ubiquitous, protecting user privacy has become challenging. With the increase of user privacy awareness and advent of privacy regulations and policies, there is an emerging need to implement software systems that enhance the protection of personal data processing. However, existing data protection and privacy regulations provide key principles in high-level, making it difficult for software engineers to design and implement privacy-aware systems. In this paper, we develop a taxonomy that provides a comprehensive set of privacy requirements based on four well-established personal data protection regulations and privacy frameworks, the General Data Protection Regulation (GDPR), ISO/IEC 29100, Thailand Personal Data Protection Act (Thailand PDPA) and Asia-Pacific Economic Cooperation (APEC) privacy framework. These requirements are extracted, refined and classified into a level that can be used to map with issue reports. We have also performed a study on how two large open-source software projects (Google Chrome and Moodle) address the privacy requirements in our taxonomy through mining their issue reports. The paper discusses how the collected issues were classified, and presents the findings and insights generated from our study. Mining and classifying privacy requirements in issue reports can help organisations be aware of their state of compliance by identifying privacy requirements that have not been addressed in their software projects. The taxonomy can also trace back to regulations, standards and frameworks that the software projects have not complied with based on the identified privacy requirements.


\end{abstract}

}

\maketitle


%
\IEEEpeerreviewmaketitle

\input{sections/1-intro}

\input{sections/2-related-work}

\input{sections/3-taxonomy-update}

\input{sections/4-mining-new}

\input{sections/5-analysis}

\input{sections/6-discussion}

\input{sections/7-conclusions}

\balance

\ifCLASSOPTIONcaptionsoff
  \newpage
\fi



\bibliographystyle{IEEEtranN}
\bibliography{TSE,ICSE2020_kookai_ref}
\end{document}

%% file: sections/1-intro.tex
\IEEEraisesectionheading{\section{Introduction}}


Software applications have become an integral part of our society. Digital footprints are collected as people are browsing the Internet or using various software applications such as for social networking, working, studying and leisure activities. Physical footprints are also collected through software systems such as surveillance cameras, face recognition apps, IoT sensors and GPS devices even when people are ``offline'' doing their normal life activities. Zettabytes of those data are collected and processed for various purposes, including extracting and using personal data, and forming behavioural profiles of individuals. This poses serious threats to our privacy and the protection of our personal sphere of life -- the cornerstone of human rights and values.

The recent advent of privacy legislations, policies and standards (e.g. the European's General Data Protection Regulation \cite{OfficeJournaloftheEuropeanUnion;2016} or the ISO/IEC standard for privacy framework in information technology \cite{ISO/IEC2011}) aims to mitigate those threats of privacy invasion. A range of frameworks (e.g. privacy by design) and privacy engineering methodologies have also emerged to help design and develop software systems that provide acceptable levels of privacy and meet privacy regulations \cite{Ayala-Rivera2018}, \cite{Solove}, \cite{Kalloniatis2008}, \cite{Deng2011}, \cite{Heurix2015}, \cite{Perera2016}, \cite{Aljeraisy2020}. However, those methodologies provide only high-level principles and guidelines, leaving a big gap for software engineers to fill in designing and implementing privacy-aware software systems \cite{Gurses2011}. Software engineers often face challenges when navigating through those regulations and policies to understand and implement them in software systems \cite{Ayala-Rivera2018}, \cite{Aljeraisy2020}, \cite{Senarath2018b}, \cite{Bednar2019}.

In 1995, \citeauthor{Cavoukian2009} initially developed a reference framework called \textit{Privacy by Design (PbD)} aiming to emphasise the importance of privacy in the engineering processes of information systems \cite{Cavoukian2009}. The framework describes seven principles that provide conceptual characteristics of privacy elements related to the processing of personal data. However, this framework is often criticised as being too broad and vague for actual implementation \cite{Rest2014}. Recently, \textit{privacy engineering} is an emerging field which addresses privacy concerns in the development of sociotechnical systems \cite{Gurses2016}. Unfortunately, the integration of privacy solutions into practices has faced several challenges and limitations due to the design of organisational processes \cite{Spiekermann2012}, \cite{Mikkelsen2019}, legacy IT systems \cite{Capgemini2019}, implementation cost \cite{Capgemini2019} and development practices \cite{Bednar2019}, \cite{Hadar2018}. Organisations have been collecting personal data of their customers for various business purposes. Cyberattacks often target at obtaining this data. CSO Online reported the fourteen biggest data breaches of the 21st century that affected 3.5 billion people \cite{Swinhoe2020}. The cases occurred with the world's top software applications, for examples, Adobe, Canva, eBay, LinkedIn and Yahoo.
 
Hence, there is an emerging need to translate complex privacy concerns set out in regulations and standards into requirements that are to be implemented in software applications. Such privacy requirements need to be refined into a level that emphasises the functionalities in software systems and can be later used to map with issue reports. Previous work has involved eliciting privacy requirements, but they are specific to a certain application domain such as e-commerce applications (e.g. \cite{Anton2002}) or healthcare websites (e.g. \cite{Antn2004}). More recent studies (e.g. \cite{Anthonysamy2017} and \cite{Guarda2009}) revealed an urgent need for a reference taxonomy of privacy requirements that are based on well-established regulations and standards such as GDPR and ISO/IEC 29100.

This taxonomy would be useful to understand how privacy requirements are explicitly addressed in existing software projects, and also serve as a basis for developing future privacy-aware software applications. Most of today's software projects follow an agile, issue-driven style in which feature requests, functionality implementations and all other project tasks are usually recorded as issues (e.g. JIRA issues\footnote{https://www.atlassian.com/software/jira}) \cite{Choetkiertikul}. In those projects, issue reports essentially contain important, albeit implicit, information about the requirements of a software, in the form of either new requirements (i.e. feature requests), change requests for existing requirements (i.e. improvements) or reporting requirements not being properly met (i.e. bugs) \cite{Choetkiertikul}, \cite{MoodleTracker}, \cite{MoodleFeature}, \cite{Merten2015}, \cite{Merten2016}. A software project consists of past issues that have been closed, ongoing issues that the team are working on, and new issues that have just been created. Through a study of those issues in the project, we can understand how the software team has implemented privacy requirements recorded in the issue tracking system (ITS) in order to address relevant privacy needs and concerns of stakeholders. This paper provides the following contributions:

\begin{enumerate}[leftmargin=0.5cm]
  \item We developed a comprehensive taxonomy of privacy requirements for software systems by extracting and refining requirements from the widely-adopted GDPR and ISO/IEC 29100 privacy framework as well as the newly developed Thailand PDPA and the region-specific APEC privacy framework. We followed a grounded theory process adapted from the Goal-Based Requirements Analysis Method (GBRAM) \cite{Antn2004}, \cite{Anton1996} to develop this taxonomy. The taxonomy consists of 7 categories and 71 privacy requirement types. To the best of our knowledge, this is the \emph{first} taxonomy of privacy requirements that are well grounded in standardised privacy regulations and frameworks.

  \item We mined the issue reports of two large-scale software projects, Google Chrome and Moodle (each has tens of thousands of issues) to extract 1,374 privacy-related issues. We classified all of those issues into the privacy requirements of our taxonomy. The classification was performed by multiple coders through multiple rounds of training sessions, inter-rater reliability assessments and disagreement resolution sessions. This resulted in a reliable dataset for the research community to perform future research in this timely, important topic of software engineering such as automated classification of privacy issue reports.

 \item We studied how the privacy requirements in our taxonomy were addressed in Google Chrome and Moodle issue reports. We found 2,307 occurrences of the privacy requirements in our taxonomy were covered, most of which related to the lawfulness, fairness and transparency privacy goal category. In addition, we found that allowing the erasure of personal data is a top concern reported in the Chrome and Moodle issue reports, while none of the privacy requirements related to the management perspective of data controllers was recorded in the issue tracking system of both projects. We also discovered that privacy and non-privacy issues were treated differently in terms of resolution time and developers' engagement in both projects.


\end{enumerate}

A full replication package containing all the artifacts and datasets produced by our studies are made publicly available at \cite{reppkg-pridp}. The remainder of this paper is structured as follows. Section \ref{sec2} provides related existing work on privacy requirements engineering. A privacy requirements taxonomy and the methodology used to build it are presented in Section \ref{sec3}. Section \ref{sec4} presents our study of how Chrome and Moodle issue reports address the privacy requirements in our taxonomy. The findings and insights generated from the study are discussed in Section \ref{results}. Section \ref{threats} discusses the threats to validity. Finally, we conclude and discuss future work in Section \ref{sec7}.

%% file: sections/2-related-work.tex
\section{Related work} \label{sec2}

The protection of personal data and privacy of users have attracted significant attention in recent years. Data protection regulations and privacy frameworks have been established to guide the development of software and information systems. \emph{However, it is challenging for software engineers to translate legal statements stated in those regulations into specific privacy requirements for software systems \cite{Ayala-Rivera2018}, \cite{Aljeraisy2020}}. Several studies have investigated the approaches, tools and techniques developed for managing privacy requirements \cite{Anthonysamy2017}. \citeauthor{Guarda2009} \cite{Guarda2009} showed that the traditional RE frameworks lack of fundamental concept specific to privacy which cannot be used to capture privacy-related legal requirements in software systems. Therefore, \citeauthor{Guarda2009} have developed a reference methodology that can be tailored to design privacy-aware systems.

\citeauthor{Beckers2012} \cite{Beckers2012} proposed a conceptual framework to compare privacy requirements engineering approaches on requirements elicitation and notion representation. The approaches include LINDDUN, PriS and the framework for privacy-friendly system design approaches. The LINDDUN method elicits privacy requirements from privacy threats \cite{Deng2011}. The method first developed nine categories of privacy threats: linkability, identifiability, non-repudiation, detectability, information disclosure, content unawareness and policy/consent non-compliance. Those threat categories are then mapped to the elements in data flow diagrams (DFDs) (i.e. entity, data flow, data store and process). The method further develops a catalogue of threat patterns (i.e. a threat tree) that represents common attacks concerned in each threat category and each element in the DFD. The patterns in the threat tree are mapped to the elements in DFD defined by misuse cases to identify privacy requirements. In the final step, the method recommends privacy-enhancing solutions that fulfill the identified requirements.

\citeauthor{Kalloniatis2008} \cite{Kalloniatis2008} proposed a PriS method to elicit privacy requirements in the software design process. Privacy requirements in this study are modelled as a type of organisational goals that needs to be achieved in a specific application. The method first identifies privacy concerns in that specific application based on the eight pre-defined privacy goals (e.g. authentication and unlinkability). The second step is to analyse impact of privacy goals on relevant processes. This impact may introduce new goals or adjustment of existing goals. After having a set of privacy-related processes, the method models those processes based on the pre-defined privacy-process patterns of each privacy goal. The final step then suggests appropriate techniques for implementing those identified processes. Comparing with our study, LINDDUN and PriS methods do not elicit requirements from privacy and data protection regulations and frameworks, and they employed different requirements elicitation approaches. The taxonomies defined in their work are selected from a set of privacy properties identified in previous studies  \cite{Solove}, \cite{George2007}, \cite{Solove2008}, \cite{anon_terminology}, \cite{BritishStandardsInstitute2000}, whereas our taxonomy is formed after eliciting requirements.

\citeauthor{Spiekermann2009} \cite{Spiekermann2009} have developed a privacy-friendly system design framework to preserve privacy concerns in software development. The framework consists of two approaches: privacy-by-policy and privacy-by-architecture. Privacy-by-policy adopts the principles in Fair Information Practices (FIPs) developed by Organisation of Economic Co-Operation and Development (OECD) \cite{OECD2013} to guide on addressing user privacy in software systems. The principles focus on providing notice and awareness to users on how their personal data will be processed. Privacy-by-architecture promotes organisations to store only necessary personal data in their systems. The authors further provide a guideline for selecting privacy protection approaches that match with system characteristics.


\citeauthor{Meis} \cite{Meis} proposed a taxonomy of transparency requirements to support software engineers in identifying relevant requirements. The requirements were derived from a draft version of GDPR and ISO/IEC 29100. The requirements formulation process includes identifying verbs and related nouns, refining identified requirements and structuring the requirements into a taxonomy of transparency requirements. This approach was later used to derive intervenability requirements from the selected articles and principles in the draft of GDPR and ISO/IEC 29100 respectively \cite{Meis2016}. These intervenability requirements were then linked to the related transparency requirements proposed earlier in \cite{Meis}. However, these studies emphasise on privacy goal transparency and intervenability in software development.

\citeauthor{Breaux2014} \cite{Breaux2014} discussed an interesting idea for software engineers on balancing between business goals and privacy requirements as both elements are essential when designing software. The business goals are normally addressed first as they drive primary activities of software applications. However, these also introduce privacy risks in order to satisfy those goals in software. The author suggests the engineers to link business requirements to privacy risks and consider ways that users could be involved in reducing privacy exposures (e.g. personalise their own level of privacy). 

Several work identified the challenges faced by software engineers when embedding privacy into software systems. First, the developers have limited knowledge and understanding of the concept of information privacy, privacy-preserving technologies and privacy practices \cite{Senarath2018b}, \cite{Hadar2018}, \cite{Senarath2018}. They interpret privacy concerns as a sub-category of security; this potentially misguides their perceptions when treating privacy in software systems \cite{Hadar2018}, \cite{Senarath2018}. Moreover, they lack evaluation criteria to assess whether they meet privacy expectations in their design \cite{Senarath2018b}. 

Second, organisation's policy also has a strong influence on developers' privacy-preserving behaviour \cite{Hadar2018}, \cite{Senarath2018}. The developers likely respond to privacy concerns depending on what they are told to do. Thus, privacy decisions are not driven by law or engineering guidelines and solutions \cite{Hadar2018}. In addition, the developers do not give priority to privacy requirements because they are not functional requirements or they are inconsistent with client/business requirements \cite{Senarath2018b}. 

Finally, privacy requirements are varied across countries/regions depending on different user expectations \cite{Senarath2018}, \cite{Sheth2014} and national legal enforcement (e.g. GDPR for European Union, US Privacy Act of 1974). When the privacy expectations between users and software developers do not match, it is challenging to preserve user privacy in software systems. Hence, the developers require privacy guidelines or evaluation criteria to assess whether they meet these expectations \cite{Senarath2018b}. 

Data protection and privacy laws are independently designed and enforced for specific areas, which could range from a state (e.g. California), a country (e.g. Australia) or a region (Europe). Any organisations that meet the conditions of these laws must comply. However, the developers lack guidance and also experience difficulties in understanding and extracting such privacy requirements from those required laws \cite{Ayala-Rivera2018}, \cite{Aljeraisy2020}. Several studies have been calling for frameworks and methodologies to support software engineers in designing and developing privacy-aware software systems \cite{Gurses2011}, \cite{Senarath2018b}, \cite{Sheth2014}, \cite{Birnhack2014}. 

Some existing work have aimed to identify privacy requirements and construct a privacy requirement taxonomy from privacy policies. \citeauthor{Anton2002} proposed a privacy goal taxonomy based on website privacy policies \cite{Anton2002}. The taxonomy was constructed by applying goal identification and refinement strategies based on the Goal-Based Requirements Analysis Method (GBRAM) \cite{Anton1996} to extract goals and requirements from 24 Internet privacy policies from e-commerce industries. This methodology was further used to develop the taxonomy of privacy requirements from 23 Internet health care Web sites privacy policies \cite{Antn2004}. They follow a goal mining process with heuristics to analyse and refine goals and scenarios from those privacy policies. We adapted this approach to construct the taxonomy presented in this paper.

Several work have also investigated how the introduction of GDPR  has changed the way that organisations comply with the privacy protection regulations and standards. \citeauthor{Linden2020} \cite{Linden2020} found that the online privacy policies of organisations in EU countries have improved their appearance, presentation and details. They also classified statements in privacy policies into a set of categories developed in \cite{Wilson2016}. Privacy concerns have also emerged in the area of Big Data analytics.  \citeauthor{Gruschka2018}  \cite{Gruschka2018} explored two case studies of Big Data research projects to investigate on how legal privacy requirements can be met. The study identified a set of key processes related to sensitive personal information, and proposed privacy-preserving techniques to achieve GDPR compliance. However, the study did not provide a systematic approach to link the identified processes to appropriate techniques.

The following work discusses privacy requirements elicitation from regulations and privacy policies for compliance checking in software systems. \citeauthor{Muller2019} \cite{Muller2019} developed a dataset consisting of statements in organisations' privacy policies. Those statements were annotated with only five privacy requirements extracted from GDPR. This dataset assisted the development of automated tools for checking GDPR compliance of an organisation's privacy policies. \citeauthor{Torre} \cite{Torre} proposed a different approach to check if an organisation's  privacy  policies  comply  with  GDPR. They first developed a conceptual model using hypothesis coding to specify metadata types that exist in the statements of selected GDPR articles. They created the dependencies between the metadata types to ensure the proper completeness checking. They then employed Natural Language Processing and Machine Learning techniques to automatically classify the information content in privacy policies from the fund domain and determine the extent to which the content satisfies a certain completeness criterion. However, the metadata types and their dependencies in the model are limited as several key articles in the GDPR were not considered (e.g. data subjects' rights and security of processing). In addition, additional criteria for completeness checking should be considered since the presence or absence of those metadata may not be sufficient to determine the completeness of statements in privacy policies against GDPR.

\citeauthor{Ayala-Rivera2018} \cite{Ayala-Rivera2018} proposed an approach to map GDPR data protection obligations with privacy controls derived from ISO/IEC standards. Those links help elicit the solution requirements that should be implemented in a software application. However, their study focused and validated only two articles in GDPR (i.e. Article 5 and 25).  \citeauthor{Torre2019} \cite{Torre2019} explored the use of models to assist GDPR compliance checking. They developed a UML class diagram representation of GDPR. This UML model is enriched with invariants expressed in the Object Constraint Language to captures GDPR rules.

Several work have investigated consistency and completeness verification for privacy policies. \citeauthor{Breaux2014a} \cite{Breaux2014a} translated a small subset of commonly found privacy requirements into description logic (a formal knowledge representation language). This enables the use of formal tools to detect conflicting privacy requirements within a policy. \citeauthor{Bhatia2019} \cite{Bhatia2019} proposed to represent data practice descriptions as semantic frames to identify incompleteness in privacy policies in organisations. They analysed 15 policies and identified 17 semantic roles associated with four categories of data action (collection, retention, usage and transfer) to express the incompleteness in data action context. Compared to our study, the settings of developing semantic roles are similar to the process we used to extract privacy requirements from the statements in the selected privacy regulations and frameworks. We can extend the use of semantic roles to categorise the scenarios of actions stated in those regulations and frameworks.

\citeauthor{Pandit2019} \cite{Pandit2019} developed the Data Privacy Vocabulary (DPV) which is an ontology of generic concepts and relationships of the components identified in GDPR. This ontology is used for completeness and compliance checking in privacy policies, consent receipts and records of personal data handling. However, their taxonomy does not cover any specific software requirements. In addition, the taxonomy addresses only six articles in the GDPR, while our work covers more articles in the GDPR and other well known data protection regulations and privacy frameworks.

Particularly focusing on the work related to GDPR, the European Commission has funded the GDPR cluster projects to help tackle the GDPR implementation challenges faced by organisations \cite{EUcluster2020}. Those projects have developed both organisational and technical techniques to facilitate the implementation. They have addressed different challenges complying with GDPR in software development activities (e.g. planning, design, development, operation and deployment). They also provide solutions to the identified challenges. \citeauthor{EUcluster2020} \cite{EUcluster2020} has summarised the solutions proposed by these projects. 	

The Business Process Re-engineering and functional toolkit for GDPR compliance project\footnote{https://www.bpr4gdpr.eu/} (BPR4GDPR) provides an approach and a toolkit to support end-to-end GDPR-complaint business processes, particularly for small and medium-sized enterprises (SMEs) \cite{BPR4GDPR}. The deliverables include the policy-based access and usage control framework, specification of workflow models and tools for cryptography, access management and enforcement of data subjects' rights \cite{EUcluster2020}.

As there is neither specific methods, techniques or tools to evaluate the GDPR readiness level in organisations nor privacy governance guideline available, the Data Privacy Governance for Supporting GDPR project\footnote{https://www.defendproject.eu/} (DEFeND) introduces a platform to assist in complying with the GDPR \cite{DEFEND}. The platform supports organisations in designing and developing tailored solution that covers different aspects of GDPR. It also provides methods and techniques to handle personal data and consent management as well as data protection mechanisms in software systems.

SMOOTH project\footnote{https://smoothplatform.eu/about-smooth-project/} aims to provide support for micro enterprises to adopt GDPR as a cloud-based platform. The solution helps assess the compliance of the enterprises and identify the basic requirements that need to be satisfied to comply with GDPR \cite{SMOOTH}. After the enterprise provides the information related to its current data protection management through the questionnaire in the platform, the platform will generate a compliance report with appropriate guidance to resolve non-compliance in the enterprise.

The Privacy and Data Protection 4 Engineering (PDP4E) project\footnote{https://www.pdp4e-project.eu/} develops a set of systematic methods and tools to address the following disciplines in software development life cycle: risks assessment, requirements engineering, model-driven design and systems assurance \cite{PDP4E}. Particularly focusing on requirements engineering, the project has introduced PDP-ReqLite, an approach to elicit privacy and data protection meta-requirements using a lightweight version of problem frames \cite{Ferreyra2020}.

The PlAtform for PrivAcY preserving data analytics (PAPAYA) project \footnote{https://www.papaya-project.eu/} develops a platform that addresses privacy preservation in data analytics. Finally, the Protection and control of Secured Information by means of a privacy enhanced Dashboard (PoSeID-on) project \footnote{https://www.poseidon-h2020.eu/} provides a solution allowing end users to manage their personal data and safeguarding the rights of data subjects \cite{POSEIDON}. The project also ensures GDPR-compliant data management and processing for organisations. The project develops a dashboard presenting the summary of their personal data (e.g. the personal data that is allowed to be shared and history of personal data transactions). The end users will be notified when data processors require the permissions to use their personal data. The data processors can also check what personal data is shared to them.

%% file: sections/3-taxonomy-update.tex
\section{A taxonomy of privacy requirements} \label{sec3}

Many countries around the world have been developing data protection and privacy legislation to strengthen their personal data and privacy protection \cite{UNCTAD2020}. These legislations are designed to provide organisations with a comprehensive benchmark to govern their personal data collection and processing as well as protect and empower individuals about their privacy and rights. Having the legislations in place seems to benefit both organisations and individuals, however many organisations have faced several challenges to comply with these legislations \cite{Capgemini2019}.

Capgemini Research Institute \cite{Capgemini2019} has produced an insightful survey on how companies have been coping with data protection and privacy compliance. They surveyed 1,100 compliance, privacy, data protection and IT executives across ten countries and eights sectors. In-depth interviews with experts in data protection and privacy regulations and practices were also conducted. The results revealed that a major challenge the businesses are facing is the alignment of existing IT systems to data protection and privacy regulations. There are top three barriers for the businesses to comply with GDPR reported in the survey. Firstly, the businesses found that aligning the legacy systems to GDPR requirements is very complex. Secondly, the GDPR requirements are too complex, which require more effort for implementation. Finally, the costs of achieving alignment with GDPR are restricted. Those challenges raise the need to develop a taxonomy of requirements from data protection and privacy regulations to support the development and compliance of privacy-aware software systems.

Our work is based on two well-established and widely-adopted regulations and privacy frameworks: GDPR and ISO/IEC 29100 and two region-specific representatives: Thailand PDPA and APEC privacy framework. GDPR is enacted to protect the individual rights of the data subjects on their personal data \cite{OfficeJournaloftheEuropeanUnion;2016}. It provides conditions, principles and definitions that need to be integrated into organisational processes and policies. These processes involve the collection, use, process, storage and dissemination of personal data of EU citizens and residents. GDPR will have a significant impact on technology platforms and data structures, data protection measures and emerging technologies implemented in organisations \cite{Li1}. The organisations failing to comply with the GDPR can be fined up to \EUR{20} million or 4\% of their previous year’s global turnover, whichever is greater. After a year of the enforcement, there are over 230 finalised cases with the total of \EUR{150} million fines so far. The largest fined case is Google Inc. in France with \EUR{50} million fines due to the lack of consent for advertisements \cite{Data}. A great number of GDPR violation cases related to the processing of personal data and data breach have been reported \cite{EuropeanCommission2019}, \cite{CNET}, \cite{PrivacyAffa}. This suggests the challenges in operationalising GDPR in developing software applications.

ISO/IEC 29100:2011 is a privacy framework which guides the processing of personally identifiable information (PII) in Information and Communication Technology systems \cite{ISO/IEC2011}. The framework defines a set of privacy principles used to handle personal data processing activities (e.g. collection, storage, use, transfer and disposal). \emph{Similarly to GDPR, those principles are high-level, making it challenging for software engineers to design and implement privacy-aware systems}. We aim to address these challenges by translating these complex statements into implementable privacy requirements for software systems.

Thailand Personal Data Protection Act (PDPA) was officially announced in May 2019 \cite{PDPA}. The regulation will come into full effect in June 2022 after several extensions due to COVID-19 disruptions. Thailand PDPA is designed to govern personal data protection and create transparency and fairness for the use of personal data. It also promotes the use of personal data for innovation under assurance and provides effective remedy from data breaches. Any organisations in Thailand that collect, use and disclose personal data must comply with the regulation. In addition, any organisations that are located outside Thailand, but sell products, provide services and/or track individuals residing in Thailand, must also comply to this regulation. We include this regulation in our study as it is a representative of newly developed and country-specific personal data protection regulation.

Asia-Pacific Economic Cooperation (APEC) privacy framework 2015 \cite{Apec2015} was published in August 2017 with the intention to establish effective privacy protections for cross-border information transfers across member countries of APEC \footnote{A list of member countries can be found at https://www.apec.org/about-us/about-apec}. This updated version was improved from the previous version published in 2005. The framework was developed based on the Organisation for Economic Co-operation and Development (OECD) guidelines. 

Prior to this work, we have conducted a thorough study on GDPR, ISO/IEC 29100, Thailand PDPA and APEC privacy framework and have found that they share many commonalities. All of the regulations, standards and frameworks provide benchmarks for privacy and data protection governance and compliance in organisations. They are common in laying out the expectations that should be met when handling personal data. They also complement each other to cover a comprehensive set of privacy-related software requirements. In addition, both GDPR and ISO/IEC 29100 define common key actors and their roles in processing personal data. The basic principles between GDPR and ISO/IEC 29100 are similar, although they are grouped differently (see Table \ref{tab:GDPR-ISO-mapping}). The requirements derived from ISO/IEC 29100 are similar or equivalent to those in GDPR. Although the terms and definitions in GDPR and ISO/IEC 29100 use different wordings, but they in fact refer to the same or similar things.

	\begin{table*}[]
	\center
	\caption{A table mapping between the GDPR, ISO/IEC 29100, Thailand PDPA and APEC framework principles.}
	\label{tab:GDPR-ISO-mapping}
	\begin{tabular}{p{3cm} p{5cm} p{4.2cm} p{3.8cm}}
		\toprule
		\textbf{GDPR principles} &
		\textbf{ISO/IEC 29100 principles} &
		\textbf{Thailand PDPA parts} &
		\textbf{APEC principles} \\ \midrule
		Lawfulness, fairness and transparency &
		\begin{tabular}[t]{@{}l@{}}Consent and choice\\ Purpose legitimacy and specification\\ Collection limitation\\ Use, retention and disclosure limitation\\ Openness, transparency and notice\\ Individual participation and access\end{tabular} &
		\begin{tabular}[t]{@{}l@{}}General provisions\\ Personal data collection\\ Use or disclosure of personal data\\ Rights of the data subject\end{tabular} &
		\begin{tabular}[t]{@{}l@{}}Preventing harms\\ Notice\\ Choice\\ Access and correction\\ Accountability\end{tabular} \\ \midrule
		Purpose limitation &
		\begin{tabular}[t]{@{}l@{}}Consent and choice\\ Purpose legitimacy and specification\\ Use, retention and disclosure limitation\\ Openness, transparency and notice\\ Accountability\end{tabular} &
		\begin{tabular}[t]{@{}l@{}}General provisions\\ Personal data collection\\ Use or disclosure of personal data\end{tabular} &
		\begin{tabular}[t]{@{}l@{}}Notice\\ Uses of personal information\\ Choice\end{tabular} \\ \midrule
		Data minimisation &
		\begin{tabular}[t]{@{}l@{}}Collection limitation\\ Data minimisation\\ Use, retention and disclosure limitation\end{tabular} &
		\begin{tabular}[t]{@{}l@{}}General provisions\\ Personal data collection\\ Rights of the data subject\end{tabular} &
		Collection limitation \\ \midrule
		Accuracy &
		\begin{tabular}[t]{@{}l@{}}Accuracy and quality\\ Individual participation and access\end{tabular} &
		Rights of the data subject &
		\begin{tabular}[t]{@{}l@{}}Collection limitation\\ Integrity of personal information\\ Access and correction\end{tabular} \\ \midrule
		Storage limitation &
		\begin{tabular}[t]{@{}l@{}}Data minimisation\\ Use, retention and disclosure limitation\\ Openness, transparency and notice\end{tabular} &
		\begin{tabular}[t]{@{}l@{}}Personal data collection\\ Rights of the data subject\end{tabular} &
		None \\ \midrule
		Integrity and confidentiality &
		Information security &
		Rights of the data subject &
		\begin{tabular}[t]{@{}l@{}}Preventing harms\\ Security safeguards\\ Accountability\end{tabular} \\ \midrule
		Accountability &
		\begin{tabular}[t]{@{}l@{}}Collection limitation\\ Accountability\\ Privacy compliance\end{tabular} &
		\begin{tabular}[t]{@{}l@{}}Use or disclosure of personal data\\ Rights of the data subject\end{tabular} &
		\begin{tabular}[t]{@{}l@{}}Preventing harms\\ Accountability\end{tabular}
	\end{tabular}
\end{table*}	

The principles in both GDPR and ISO/IEC 29100 present a set of basic guidelines to govern personal data and privacy protection covered in the regulations and standard. We have found that there are commonalities and differences across these sources. For instance, the integrity and confidentiality in GDPR and information security principle in ISO/IEC 29100 address the common concerns on having appropriate measures put in place to protect personal data and its processing. The examples of the requirements related to these principles are \emph{R60 IMPLEMENT appropriate technical and organisational measures to protect personal data} and \emph{R63 PROTECT the personal data from unauthorised access and processing}.

It is also interesting to note that requirements from a principle in ISO/IEC 29100 can be addressed in different principles in GDPR. For example, we have derived requirements \emph{R1 ALLOW the data subjects to access and review their personal data} and \emph{R50 INFORM the recipients of personal data any rectification or erasure of personal data or restriction of processing} from the individual participation and access principle in ISO/IEC 29100. Requirement R1 reflects the lawfulness, fairness and transparency principle in GDPR while requirement R50 relates to the accuracy principle in the GDPR. 

This section presents a taxonomy of privacy requirements that we have developed based on the GDPR, ISO/IEC 29100, Thailand PDPA and APEC privacy framework. We first discuss the methodology that we have followed to develop this taxonomy, and then describe the privacy requirements set out in the taxonomy.

\subsection{Methodology} \label{sec3.1}

\begin{figure}
	\centering
	\includegraphics[width=1.0\linewidth]{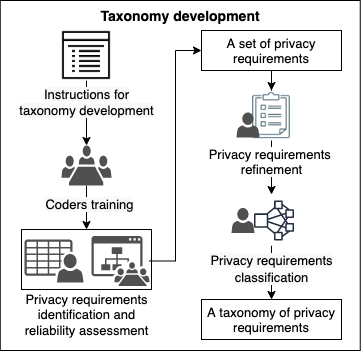}
	\caption{An overview process of privacy requirements taxonomy development}
	\label{fig:taxonomy development}
\end{figure}

We followed a content analysis process adapted from the Goal-Based Requirements Analysis Method (GBRAM) \cite{Antn2004}, \cite{Anton1996}, which is based on Grounded Theory \cite{Glaser2017}, to develop a taxonomy of privacy requirements (see Figure \ref{fig:taxonomy development}). GBRAM is a systematic method used to identify, refine and organise goals into software requirements. This process was applied to analyse goals from natural language texts in privacy policies, and convert them into software requirements. The method has been successfully applied to the analysis of e-commerce applications \cite{Anton1998}, \cite{Baumer2000} and Internet health care Web site privacy policies \cite{Antn2004}. The method consists of three main activities: goal identification, goal classification and goal refinement. Goal identification derives goals from specifications in selected sources. Each identified goal is then classified into one of the pre-defined categories in the goal classification. Finally, the goal refinement removes synonymous and redundant goals, resolves inconsistencies among the goals and operationalises the goals into requirements specification. 

We had multiple researchers (co-authors of the paper, hereby referred to as the coders) follow this process to develop the taxonomy independently, and used the Inter-Rater Reliability (IRR) assessment to validate the agreements and resolve disagreements. Those coders were given instructions and trained at the start of the process. The process consists of the following steps:

\begin{itemize}[leftmargin=0.5cm]
	\item \textbf{Privacy requirements identification:} extract requirements from written statements in the studied privacy regulations and frameworks, and structure them into a pattern (action verb, objects and object complement).

	\item \textbf{Privacy requirements refinement:} remove duplicate requirements and manage inconsistent requirements. Since the inputs were written in descriptive statements and from different sources, requirements can be redundant or inconsistent.
	
	\item \textbf{Privacy requirements classification:} classify requirements into categories based on a set of privacy goals. The privacy goals can be considered as a group of functionalities that the software systems are expected to provide.
\end{itemize}

\vspace{0.5mm}

The details of each step are described as follows.

\subsubsection{Privacy requirements identification}

This step aims to identify privacy requirements from the narrative statements in GDPR, ISO/IEC 29100 privacy framework, Thailand PDPA and APEC privacy framework. We adapted the approach in \citeauthor{Antn2004} \cite{Antn2004} and first created a range of questions to identify goals from each statement of the studied privacy regulations and frameworks. As part of our research, we have carefully gone through all 99 articles in the GDPR. Although GDPR and Thailand PDPA govern broader regulatory aspects comparing to ISO/IEC 29100 and APEC privacy frameworks such as requirements in roles assignment (e.g. data protection officer, supervisory authority, personal data protection committee), managing juridical remedies and noticing penalties, those are not software requirements. Thus, they are out of the scope of our study. 

We then applied several filters to select the articles that address software requirements in GDPR and include them in our study. We selected nineteen articles that address the rights of individuals and cover the key principles of the GDPR (i.e. Articles 6-7, 12-22, 25, 29-30, 32-34). We also excluded the articles that deal with special conditions (e.g. Article 8 child's personal data and Article 9 special categories of personal data) since we aim to keep our taxonomy as generalisable as possible for software systems in different sectors. International transfers are presented in Articles 13(1)(f), 14(1)(f) and 15(2), which we have considered in our study. We did not consider Chapter V (Articles 44 - 50) since it focuses on legal administrative perspective for international border transfer (e.g. the provisions determining conditions for international transfer and international cooperation for the protection of personal data) rather than the software application level. Thus, we did not analyse them as software requirements in our study. For the ISO/IEC 29100 privacy framework, all of the contents were explored.

Thailand PDPA consists of 7 chapters, 96 articles and 7 rights of data subjects. We have analysed 16 articles in Chapter 2 (Personal data protection) and Chapter 3 (Rights of data subjects) which are the key chapters providing guidelines to govern personal data and privacy protection. Chapter 2 consists of three parts: i) general provisions, ii) personal data collection and iii) use or disclosure of personal data. Other chapters detailing the scope of use, definitions, assignment of Personal Data Protection Committee and supervisory authority, complaints and penalties are not included in the taxonomy development process as they are not related to software requirements. In the APEC privacy framework, there are four parts containing 72 points: i) preamble, ii) scope, iii) APEC information privacy principles and iv) implementation. We have analysed the third part to derive privacy requirements in the taxonomy development process. The third part includes nine information privacy principles: i) preventing harms, ii) notice, iii) collection limitation, iv) uses of personal information, v) choice, vi) integrity of personal information, vii) security safeguards, viii) access and correction and ix) accountability. The other parts define the objectives, scope and organisational-level guidelines and are not related to software requirements. Based on the scope defined above, we shortlisted 149 statements in GDPR, 63 in ISO/IEC 29100, 101 in Thailand PDPA and 74 in the APEC privacy framework to be explored.

Once we selected the articles and sections to work on, we went through all the sentences in those articles and sections. We analysed each sentence using a set of pre-defined questions to identify relevant actions, involved/affected parties or objects and target results. A sentence may cover more than one requirement. There are three key components for a requirement: 1) actions, 2) involved/affected parties or objects, and 3) target results. The detailed steps of privacy requirements identification process are explained below:

\begin{enumerate}[leftmargin=*]
	
	\item \textbf{Identifying actions:} We ask \textit{``Which action should be provided based on this statement?''} to identify the action associated with a requirement. Some examples of the action verbs used in the collected statements are: \textbf{ALLOW, ARCHIVE, COLLECT, ERASE, IMPLEMENT, INFORM, MAINTAIN, NOTIFY, OBTAIN, PRESENT, PROTECT, PROVIDE, REQUEST, SHOW, STORE, TRANSMIT and USE}.
	
	\item \textbf{Determining involved/affected parties or objects:} After an action is identified, it is necessary to determine the object(s) of the action. The output from the second step in the privacy requirements identification process can be either involved/affected parties or objects. The involved/affected parties can be any persons or stakeholders such as data subjects, data processors, supervisory authorities and third parties. The question used to identify the involved/affected parties is ``Who is involved/affected from the statement?''. However, the objects are things that are created, processed or done by the actions specified in the statements (e.g. consent, preferences, personal data and functions). These objects are identified by asking ``What has to be created/done from the identified action?''. If a statement refers to a data source or a data repository, it can be extracted as an object, while a data recipient can be extracted as the involved/affected party.
	
	\item \textbf{Considering the target result(s):} The target results refer to a goal that a statement aims to achieve. They can be identified by asking \textit{``What should be achieved based on the action of that statement?''} For example, Article 13(1)(c) in GDPR states ``..., the controller shall, at the time when personal data are obtained, provide the data subject with the purposes of the processing for which the personal data are intended as well as the legal basis for the processing;''. The goal in this statement is asking to provide the data subject with the purposes of processing. Hence, the purposes of the processing is a target result that the action verb \textit{PROVIDE} aims to achieve.
	
	\item \textbf{Structuring into a privacy requirement pattern:} The derived privacy requirement is coded in the format of action verb, followed by involved/affected parties or objects and target results.
	
\end{enumerate}

In our study, the statements in the regulations, standards and frameworks are written in natural language, and they contain goals that shall be complied or satisfied. We derived those privacy goals and then constructed them as software privacy requirements. Thus, this method meets our objective in terms of the inputs that we have and outputs that we would like to achieve. There are three essential components in a statement in the privacy regulations, standards and frameworks: what to do (i.e. an action), to whom/what they apply (i.e. an affected/involved party or an object) and things they aim to achieve (i.e. an object complement). Our requirements identification process captured all of those three parts. Several studies also constructed privacy requirements by including these three components from the statements in the data protection regulations and standard \cite{Meis} and privacy policies \cite{Antn2004}.

The following examples illustrate these steps. A statement in the GDPR states ``... the controller shall, at the time when personal data are obtained, provide the data subject the identity and the contact details of the controller and, where applicable, of the controller's representative''. From this statement, we identify \textit{`PROVIDE'} as the action that the controller shall act. We then consider what should be provided by the controller, and that was \textit{`the identity and the contact details of the controller or the controller's representative'}. We determine the object responding to \textit{`to whom the identity and contact details of the controller or the controller's representative should be provided'}, and that is the data subjects. All three components formulate a privacy requirement as \textit{\textbf{PROVIDE} the data subjects with the identity and contact details of a controller/controller's representative (R22)}.

Another example is more complex than the previous one. In the GDPR, removing personal data is recommended in several ways such as: the data has been unlawfully processed; or the data subjects would like to erase their personal data themselves; or the system must erase personal data when the data subjects object to the processing; or the system must erase personal data when the data subjects withdraw consent; or the system must erase personal data when it is not necessary for the specified purpose(s); or the system must erase personal data when the purpose(s) for the processing has expired. We thus need to formulate different privacy requirements as they affect the ways that the functions would be provided to users in a system. In this example, the derived requirements are: \textit{\textbf{ERASE} the personal data when it has been unlawfully processed (R7)}, \textit{\textbf{ALLOW} the data subjects to erase his/her personal data (R44)}, \textit{\textbf{ERASE} the personal data when the data subjects object to the processing (R46)}, \textit{\textbf{ERASE} the personal data when a consent is withdrawn (R47)}, \textit{\textbf{ERASE} the personal data when it is no longer necessary for the specified purpose(s) (R52)} and \textit{\textbf{ERASE} the personal data when the purpose(s) for the processing has expired (R53)}.

It is important to note that we have followed a systematic approach (GBRAM) to extract and refine requirements from narrative statements. For some statements, they are straightforward since privacy requirements can be directly derived from them. However, we have restructured and refined some privacy requirements to emphasise the functionalities that should be provided by software systems. For example, Article 7(1) in the GDPR states ``Where processing is based on consent, the controller shall be able to demonstrate that the data subject has consented to processing his or her personal data''. From this statement, we derive requirement \emph{R17 - SHOW the relevant stakeholders the consent given by the data subjects to process their personal data}.

The lawfulness of processing is one of the key principles for protecting privacy of data subjects. Based on the GDPR Article 6, the processing of personal data will be lawful if the data subjects give consent \emph{or} the processing is required for other legal conditions, such as the processing is necessary for the performance of contract, compliance with a legal obligation and protecting vital interests. R35 in the taxonomy addresses the requirement of obtaining consent from the data subjects for the processing. For other legal conditions, the data controllers must inform the data subjects if they need to process personal data under those conditions. Requirements R38 and R39 in the taxonomy cover these scenarios.

After lawfully obtaining personal data, the data controllers must provide the data subjects with mechanisms to execute their individual rights in the system. All the privacy requirements related to the rights of data subjects are covered in our taxonomy (i.e., the right to be informed (R12, R24, R26, R30, R31 and R50), the right of access (R1), the right to rectification (R45), the right to erasure (R44), the right to restriction of processing (R4), the right to data portability (R33), the right to object (R3), and the rights in relation to automated decision making and profiling (R21)).

Our taxonomy has a number of requirements (e.g., R35, R39 and R60) that can be triggered when the software is dealing with special category data/more sensitive data. These requirements related to GDPR Art. 9 which discusses the lawful processing of special categories of personal data. According to GDPR Art. 9, in addition to consent, there are other conditions for a lawful processing of special categories of personal data (e.g. necessary for protecting vital interests, substantial public interests and the purposes of preventive or occupational medicine). Requirement R35 covers the consent condition. R39 and R60 in our taxonomy address the remaining conditions specified in GDPR Art. 9. R39 requires the data controllers, if not obtaining the consent, must provide the data subjects the purpose(s) of the processing of special categories of personal data to the data subjects. The controllers also require to protect those personal data with appropriate measures (R60).

The proposed taxonomy would benefit software developers, data controllers, data processors and DPOs. However, there are several reasons we did not include the articles related to the DPOs in our study. Firstly, the articles related to DPOs in GDPR mainly focus on their duties, tasks and responsibilities (i.e. Articles 37 - 39). These include how the DPOs govern, supervise and cooperate with the data controllers, the data processors, the supervisory authority and other stakeholders regarding the processing of personal data. In addition, the DPOs must ensure that the processing of personal data carried out in organisations complies with data protection regulations or other relevant laws. Thus, the DPOs requirements are related to governance aspect rather than software requirements aspect.
	
Secondly, the main role that directly determines the activities related to the processing of personal data is the data controllers. Any concerns about the processing of personal data will be notified to the DPOs by the data controllers, and sometimes the data processors. This makes the data controllers the key stakeholder in governing how personal data is processed and how the processing activities should be done in software development level. Finally, we have not found any DPOs requirements reported in issue reports in our study. This finding implies that the DPOs requirements were not reflected as software requirements in this context.

Our taxonomy covers three key privacy requirements related to international data transfer as follows: i) the data subjects must be informed about the transfer of their personal data to a third country or an international organisation (R9); ii) the personal data must be appropriately protected (R60) and iii) the transfer of personal data must comply with local requirements (R65).

We note that an article, a section or a statement can lead to the identification of more than one requirements. For example, we derived 17 privacy requirements (i.e. R1-R4, R6, R9, R20-22, R27, R29, R34, R37, R39, R44-R45 and R55) from the Article 13 in GDPR\footnote{See the file \emph{Privacy-requirements-references} in the replication package for more details \cite{reppkg-pridp}.}. The following example demonstrates two privacy requirements that were derived from a statement. A statement in Section 23-6 in Thailand PDPA states ``In collecting the Personal Data, the Data Controller shall inform the data subject, ... (5) information, address and the contact channel details of the Data Controller, where applicable, of the Data Controller's representative or data protection officer;'', we derived two requirements from this statement which are \textit{R20 PROVIDE the data subjects with the contact details of a data protection officer (DPO)} and \textit{R22 PROVIDE the data subjects with the identity and contact details of a controller/controller's representative}.

\textbf{Reliability assessment}: Three human coders, who are the co-authors of the paper, have independently followed the above process to identify privacy requirements from the GDPR and ISO/IEC 29100 privacy framework. All three coders had substantial software engineering background and at least 1 year of experience with data protection regulations and policies. The first author prepared the materials and detailed instructions\footnote{These are included in the replication package \cite{reppkg-pridp}.} for the process. The instructions were provided to all the coders before they started the identification process. A 1-hour training session was also held to explain the process, clarify ambiguities and define expected outputs. The coders also went through a few examples together to fine tune the understanding.

All the coders were provided with a form to record their results of each step. The form was pre-filled with 149 statements extracted from the GDPR and 63 statements from the ISO/IEC 29100 privacy framework. If a coder considers a statement as a privacy requirement, they need to identify the relevant components and structure it following the privacy requirement patterns above. Otherwise, they leave it blank. Initially, the three coders each respectively identified 100, 95 and 97 requirements from GDPR, and 36, 36 and 37 requirements from ISO/IEC 29100.

Since the requirements identified by the coders could be different, we used the Kappa statistic (also known as Kappa coefficient) to measure the inter-rater reliability between the coders \cite{Viera2005}. The Kappa statistic ranges from -1 to 1, where 1 is perfect agreement and -1 is strong disagreement \cite{Viera2005}. There are several types of the Kappa statistics which suit different study settings \cite{Hallgren}. For this study, the Fleiss' Kappa was used as we had three coders coding the same datasets \cite{Fleiss1971}. The Kappa values were 0.8025 for GDPR and 0.7182 for ISO/IEC 29100, suggesting a substantial agreement level amongst all the coders \cite{Landis1977}. All the coders agreed that there were 43 and 20 statements from the GDPR and ISO/IEC 29100 respectively that are not privacy requirements. There were 20 GDPR statements (and 13 for ISO/IEC 29100) that the coders did not agree upon. Hence, a meeting session was held between the coders to discuss and resolve disagreements. 

The statements in Thailand PDPA and APEC privacy framework were identified by one of the coders using the same methodology performed with the GDPR and ISO/IEC 29100. All the shortlisted statements in Thailand PDPA and APEC privacy framework were derived in this step. This brought the total number of privacy requirements obtained in this step to 249 (116 from GDPR, 33 from ISO/IEC 29100 and 55 from Thailand PDPA and 45 APEC privacy framework).

We used only one coder for the Thailand PDPA and APEC privacy framework because the statements are less complex than the GDPR and ISO/IEC 29100. The provisions in Thailand PDPA were largely written based on the GDPR \cite{Kateifides}, thus we did not require additional process or special needs as they share many commonalities. Similarly, many principles in the APEC privacy framework are similar to ISO/IEC 29100. We therefore decided that one coder would be sufficient. The coder, who was responsible for the Thailand PDPA and APEC privacy framework, was the main coordinator and also participated in the privacy requirements identification process for GDPR and ISO/IEC 29100.

\subsubsection{Privacy requirements refinement} \label{refinement}

Requirements extracted either from the same or different sources (e.g. GDPR and ISO/IEC 29100) can be similar, redundant or inconsistent. In this step, we identify those similar and duplicate requirements, and merge them into one single requirement. In case that the requirements are inconsistent, we perform further investigation and report for notice.

To identify and merge similar requirements, we first place those similar requirements into the same group. These requirements tend to achieve the same goal and have the same involved/affected parties or objects. We then determine the action and target result for the final merged requirement based on the following rules:
\begin{enumerate}
	\item If the action verbs in the requirements are the same, we retain that action for the final merged requirement.
	\item If the actions are different, we consider the action verb based on the following:
	\begin{itemize}
		\item Use \emph{ALLOW} if a requirement relates to data subject's ability to invoke his/her rights.
		\item Use \emph{PROVIDE} if a requirement aims to give information to stakeholders.
		\item Use \emph{OBTAIN} if a requirement aims to get a consent or permission from stakeholders.
		\item Use \emph{PRESENT} if a requirement aims to display options or choices to stakeholders. This action verb requires responses from the stakeholders (e.g. displaying toggles or radio buttons for users to select).
		\item Use \emph{SHOW} if a requirement aims to show information to stakeholders. This action verb does not require any responses from the stakeholders.
		\item Use \emph{NOTIFY} if a requirement aims to alert stakeholders.
		\item Use \emph{IMPLEMENT} if a requirement aims to build a mechanism to support an activity in a system.
		\item Use \emph{ERASE} if a requirement aims to erase data in software systems.
	\end{itemize}
	\item If the target results in the requirements are the same, we retain that target result for the final requirement.
	\item If the target results are different, we combine them together. In case they have redundant or synonymous words, we select one word from the words in the list.
\end{enumerate}	

We finally put together the action, involved/affected parties or objects and target results identified in the above steps to construct the final requirement. 

We have carefully investigated the terms and definitions used in GDPR and ISO/IEC 29100 and found that they mostly refer to similar or same things. For example, ISO/IEC 29100 defines personally identifiable information or PII as ``any information that (a) can be used to identify the PII principal to whom such information relates, or (b) is or might be directly linked to a principal''. Personal data in GDPR is defined as ``any information relating to an identified or identifiable natural person ('data subject')''. As can be seen, PII in ISO/IEC 29100 and personal data in GDPR in fact refer to the same thing, i.e. any information that can be used to identify or is linkable to a natural person.

Similarly, key terms GDPR and ISO/IEC 29100 are sometimes worded differently, however their definitions are similar. For example, \textit{`processing'} means ``any operation or set of operations which is performed on personal data or set of personal data, ...'' in GDPR and \textit{`processing of PII'} is defined as ``operation or set of operations performed upon personally identifiable information (PII)''. Another example is data controller in GDPR and PII controller in ISO/IEC 29100. Both terms refer to person that determines the purposes and means of the processing of personal data/PII. Other examples include personal data with PII, data subject with PII principal, data processor with PII processor, consent and third party. Thus, in the merging step, we use the terms from GDPR in representing our requirements in this taxonomy to avoid ambiguities.

The additional example is a PII principal in ISO/IEC 29100 and a data subject in GDPR. A PII principal in ISO/IEC 29100 is defined as ``a natural person to whom the personally identifiable information relates''. A data subject in GDPR is defined as ``an identifiable natural person is one who can be identified , directly and indirectly, in particular by reference to an identifier such as a name, ...''. As can be seen, a PII principal in ISO/IEC 29100 and a data subject in GDPR in fact refer to the same thing, i.e. a natural person who can be identified with identifiable information such as name. 

We have found that software requirements derived from GDPR, ISO/IEC 29100, Thailand PDPA and APEC privacy framework are in fact at the same level of abstraction\footnote{see the file \emph{Privacy-requirements-abstraction} in the replication package for additional sample requirements \cite{reppkg-pridp}}. \citeauthor{Meis} has also confirmed that GDPR and ISO/IEC 29100 are at the same level of abstraction \cite{Meis}. The following example demonstrates this case. We derived \textit{\textbf{PROVIDE} the data subject the recipients or categories of recipients of the personal data} from Article 13(1)(e) in GDPR, \textit{\textbf{PROVIDE} the types of persons whom the PII can be transferred} from openness, transparency and notice principle in ISO/IEC 29100, \textit{\textbf{INFORM} the data subject the categories of Persons or entities to whom the collected Personal Data may be disclosed} from Section 23 in Thailand PDPA and \textit{\textbf{PROVIDE} the types of persons or organisations to whom personal information might be disclosed} from Point 21 in APEC framework. These four requirements demonstrate that they are at the same level of abstraction and aim to achieve the same goal. They can be merged in the requirements refinement process later.

The following example demonstrates the requirements merging step. A statement in ISO/IEC 29100, ``... allow a PII principal to withdraw consent easily and free of charge ...'', derives a requirement \textit{\textbf{ALLOW} a PII principal to withdraw consent}. A statement in GDPR, ``... the controller shall ... provide the data subjects with ... the existence of the right to withdraw consent at any time ...'' gives a requirement \textit{\textbf{PROVIDE} the existence of the right to withdraw consent}. A statement in Thailand PDPA, ``The data subject may withdraw his or her consent at any time.'', gives a requirement \textit{\textbf{ALLOW} the data subject to withdraw his or her consent.} The goal of these three requirements is to let the PII principal/data subject withdraw consent, and the affected parties are PII principal and data subject. It is noted that we use the terms from GDPR for roles in our requirements (i.e. data subject, data controller, data processor and third parties). We therefore list them as similar requirements. The requirements have different actions (i.e. ALLOW and PROVIDE), we then use \emph{ALLOW} as the final action as these requirements are about data user's ability to withdraw consent. We acquire \textit{withdraw consent} as a common target result. Finally, we merge these three requirements into a single requirement: \textit{\textbf{ALLOW} the data subjects to withdraw consent (R6)}.

The duplicate requirements are the requirements that have the exact actions, involved/affected parties and target results. We represent these requirements as one requirement in the taxonomy. For example, we identify two exact requirements in the identification process, \emph{PROVIDE the data subject the categories of personal data concerned} in GDPR articles 14(b) and 15(b). We retain one requirement (i.e. R42) in the taxonomy.

Requirements are inconsistent when they appear to contradict each other in performing the same actions. The following example demonstrates the consistency between the requirements in ISO/IEC 29100 and GDPR. We identify the requirements from ISO/IEC 29100 and GDPR as ``COLLECT only necessary PII for specific purposes'' and ``COLLECT the personal data as necessary for specific purposes'', respectively. Both requirements yield that the personal data must be collected as necessary for specific purposes. They are presented in both GDPR and ISO/IEC 29100. The requirements are therefore consistent, and merged as R41 COLLECT the personal data as necessary for specific purposes.

The following example is made up for the purpose of explanation to demonstrate the inconsistency between requirements. Assuming a statement states ``Any personal data can be freely collected without specifying a specific purpose for collection'', we have derived the requirement as ``COLLECT any personal data without a specific purpose''. This requirement would contradict with requirement R41 discussed above since the former does not require a specific purpose provided, while the latter does.

We merged in total 178 similar and duplicate requirements. We did not find any inconsistent requirements. For requirements traceability, we have provided a full list of the privacy requirements with their references to the GDPR articles, ISO/IEC 29100 principles, Thailand PDPA sections and APEC framework points in the replication package \cite{reppkg-pridp}. This step resulted in a final taxonomy of 71 privacy requirements in 7 goal categories which we will discuss in detail in the next subsection.

\subsubsection{Privacy requirements classification} \label{requirements-classification}

In this step, we aim to group the privacy requirements into categories based on their goals. We applied a bottom-up approach to classify the requirements into privacy goal categories. This approach ensures that the generated categories cover and address all the requirements. The approach also allows the categories in the taxonomy to be updated when there are new privacy requirements identified in the future. For example, newly identified privacy requirements can be added to existing categories or form new categories.

The bottom-up approach consists of two steps. We first considered the privacy requirements based on their actions, objects and target results. For example, the privacy requirements with the action verb \emph{ALLOW}, the object \emph{data subjects} and the target results that are related to individual rights (e.g. access, rectify and erase) were grouped together (i.e. user participation). Similarly, the privacy requirements with the action verb \emph{PROVIDE}, the object \emph{data subjects} and the target results that are related to information for stakeholders were gathered into the same group. We kept applying this strategy to group the rest of privacy requirements. We then ended up with fifteen categories in the first step.

Next, we grouped the privacy requirements that have at least two common components either the actions, objects or target results. For example, we gathered the privacy requirements that have the same action verb \emph{PROVIDE} and the same target results that are related to information for stakeholders, but there are two different objects - data subjects and other parties that are not data subjects. We then created subcategories for each object (i.e. notice - data subjects and notice- relevant parties). All of these requirements were grouped under the notice category. We repeated this step with the rest of privacy requirements. Finally, the taxonomy consists of seven categories (some of which have sub-categories): user participation, notice, user desirability, data processing, breach, complaint/request and security. Each requirement can belong to multiple categories.

The descriptions of privacy goal categories below provide an overview and briefly illustrate the privacy requirements concerned in those categories. We note that the descriptions do not explain the full list of privacy requirements in each category.

\begin{enumerate}[leftmargin=*]
	
	\item User participation \\
	This privacy goal consists of a set of requirements for the controllers to provide the data subjects with the functionalities to invoke their individual rights relating to their personal data. The data subjects must be able to access, review, rectify, erase and verify the validity and completeness of their personal data. The controllers shall provide the data subjects their personal data when they request to obtain and reuse their personal data for their own purposes for other services. The controllers shall allow the data subjects to object to and restrict the processing of their personal data. The data subjects must be able to withdraw consent or lodge a complaint to a supervisory authority.
	
	\item Notice \\
	This privacy goal consists of two sub-categories: data subjects and relevant parties. This category has a set of requirements for the data subjects to be informed and/or notified of relevant information and individual rights related to the processing of personal data. The information includes privacy policies, procedures, practices and logic of the processing of personal data. The data subjects must be informed of the purposes of collection and processing of their personal data. The controllers must provide the contact details of responsible persons who control the processing. The data subjects must be notified when they are likely to be in risk from personal data exposures. In addition, the controllers must communicate any relevant information relating to personal data processing to intended stakeholders.
	
	\item User desirability \\
	This privacy goal consists of three sub-categories: consent, choice and preferences. This category asserts that the controllers must show a consent form to and obtain consent from the data subjects. The controllers must provide the data subjects with an option to provide their data, allow the processing or subject to a decision based on automated processing. The processing should be implemented based on user preferences expressed in their consent.
	
	\item Data processing \\
	The privacy goal addresses the processing of personal data handling from the controllers' side. The processes, which are the sub-categories in this category, include collection, use, storage, erasure, transfer and record. The controllers must handle personal data as necessary for specific purposes. The processors who process personal data must also follow the instructions from the controllers. The personal data processing activities must be recorded.
	
	\item Breach \\
	This goal category ensures that the controllers must be prepared to handle personal data breach. The relevant information about the data breach must be recorded and communicated to data subjects and relevant stakeholders.
	
	\item Complaint/Request \\
	This goal category addresses complaint and request management. User complaints and requests about their individual rights and the processing of their personal data must be processed. The actions regarding the complaints and requests must be informed. The controllers must request for relevant information to confirm the identity of data subjects when the request has been made.
	
	\item Security \\
	This privacy goal ensures that personal data and its processing are safeguarded with confidentiality, integrity and availability. The personal data must be protected with appropriate controls and security mechanisms. The security mechanisms must comply with security and data protection standards. The personal data must be accessed and used by the authorised stakeholders. The controllers must ensure the correctness and completeness of personal data. 
\end{enumerate}

When software engineers map an issue report to relevant privacy requirements, they can consider the features/concerns that the issue report mentions based on categories, then explore relevant requirements under the selected categories and sub-categories. The software engineers will know what has been missed in this issue, so they can resolve the issue accordingly. For example, an issue report reports on not allowing users to modify their personal data, the software engineer can directly observe the requirements in the user participation category as it relates to the data subject's right. He/she then identifies the relevant requirement of this issue report, which in this case is requirement R45. Similarly, if the issue mentions about consent, the engineer can access the requirements related to consent in the consent sub-category, select the relevant one(s) and resolve the issue based on the identified privacy requirements.

This proposed methodology is able to address the scenario where two requirements aiming to achieve the same goal have different actors. We can further refine relevant requirements into a parent requirement and child requirements with specific logical operators (i.e. AND and OR). For example, assume that the controller is required to obtain consent for the processing of personal data. The data controller is responsible for this task in GDPR, however this could be managed by 3rd party authority in other regulations (e.g. CCPA \cite{StateofCaliforniaDepartmentofJustice2018}). We can then refine these requirements into a parent requirement as \textit{obtain consent for the processing of personal data}. The child requirements could be expressed as \textit{the controllers shall obtain consent for the processing of personal data} and \textit{the 3rd party authority shall obtain consent for the processing of personal data}. The logical operator in this scenario is OR as software development teams can choose between one of the child requirements to implement to satisfy the parent requirement. 

If a regulation states that it requires both the controllers and the 3rd party authority to obtain consent for the processing of personal data, the logical operator in this scenario is AND, and the software development teams must implement both child requirements in their system. As a data controller is the main actor in the regulations and frameworks we analysed in this study, the above scenario did not occur in our study. Thus, we did not refer to a specific requirement in our taxonomy. \\

\subsection{Privacy Requirements Taxonomy} \label{sec3.2}

Our taxonomy consists of a comprehensive set of 71 privacy requirements classified into 7 categories. The full version of the taxonomy can be found in Appendix. We now highlight some of the important requirements in each category (see Table \ref{tab:table1}). We note that there are typically four types of roles involved in a privacy requirement: (i) data subjects who provide their personal data for processing, give consent and determine their privacy preferences; (ii) data controllers who determine what data to be collected and the purpose of personal data collection and processing; (iii) data processors who process the personal data corresponding to the specified purpose and (iv) third parties who in case receive personal data from the controllers or processors.

\begin{table}[]
		\caption{Selected privacy requirements that are referred in the paper. The full taxonomy is available in Appendix.}
		\label{tab:table1}
		\begin{tabular}{p{8.5cm}}
			\toprule 
			\textbf{Privacy requirements}\\
			\midrule 
			
			\textbf{Category 1: User participation} \\
			R1 ALLOW the data subjects to access and review their personal data \\
			R6 ALLOW the data subjects to withdraw consent \\
			R34 ALLOW the data subjects to obtain and reuse their personal data for their own purposes across different services \\
			R44 ALLOW the data subjects to erase their personal data     \\
			R45 ALLOW the data subjects to rectify their personal data \\
			
			\vspace{1mm}
			
			\textbf{Category 2: Notice} \\
			R12 INFORM the data subjects the reason(s) for not taking action on their request and the possibility of lodging a complaint \\
			R15 NOTIFY the data subjects the data breach which is likely to result in high risk \\
			R17 SHOW the relevant stakeholders the consent given by the data subjects to process their personal data  \\
			R19 PROVIDE the data subjects an option to choose whether or not to provide their personal data \\
			R22 PROVIDE the data subjects with the identity and contact details of a controller/controller's representative \\
			R26 PROVIDE the data subjects the information relating to the policies, procedures, practices and logic of the processing of personal data   \\
			R27 PROVIDE the data subjects the recipients/categories of recipients of their personal data   \\
			R30 PROVIDE the data subjects the information relating to the processing of personal data with standardised icons \\
			R38 PROVIDE the data subjects the purpose(s) of the collection of personal data      \\
			R39 PROVIDE the data subjects the purpose(s) of the processing of personal data \\
			R42 PROVIDE the data subjects the categories of personal data concerned  \\			
			R55 PROVIDE the data subjects the period/criteria used to store their data \\
			R66 NOTIFY a supervisory authority the data breach   \\
			
			\vspace{1mm}
			
			\textbf{Category 3: User desirability} \\
			R6 ALLOW the data subjects to withdraw consent \\
			R8 IMPLEMENT the data subject's preferences as expressed in his/her consent \\
			R19 PROVIDE the data subjects an option to choose whether or not to provide their personal data  \\
			R35 OBTAIN the opt-in consent for the processing of personal data for specific purposes \\
			R36 PRESENT the data subjects an option whether or not to allow the processing of personal data \\
			R47 ERASE the personal data when a consent is withdrawn \\
			
			\vspace{1mm}
						
			\textbf{Category 4: Data processing} \\			
			R7 ERASE the personal data when it has been unlawfully processed \\
			R13 MAINTAIN a record of personal data processing activities  \\
			R40 USE the personal data as necessary for specific purposes specified by the controller \\
			R41 COLLECT the personal data as necessary for specific purposes   \\
			R43 STORE the personal data as necessary for specific purposes \\
			R46 ERASE the personal data when the data subjects object to the processing \\
			R47 ERASE the personal data when a consent is withdrawn \\
			R52 ERASE the personal data when it is no longer necessary for the specified purpose(s)   \\
			R53 ERASE the personal data when the purpose for the processing has expired  \\
			
			\vspace{1mm}
			
			\textbf{Category 5: Breach} \\
			R15 NOTIFY the data subjects the data breach which is likely to result in high risk \\
			R66 NOTIFY a supervisory authority the data breach   \\
			R67 NOTIFY relevant privacy stakeholders about a data breach \\
			
			\bottomrule
			
		\end{tabular}
\end{table}

\begin{table}[]
	\label{tab:table1-2}
	\begin{tabular}{p{8.5cm}}
		\toprule 
		\textbf{Privacy requirements (Continued)}\\
		\midrule 
		
		\textbf{Category 6: Complaint/Request} \\
		R12 INFORM the data subjects the reason(s) for not taking action on their request and the possibility of lodging a complaint \\
		R31 REQUEST the data subjects the additional information necessary to confirm their identity when making a request relating to the processing of personal data \\
		
		\vspace{1mm}
		
		\textbf{Category 7: Security} \\
		R56 ALLOW the authorised stakeholders to access personal data as instructed by a controller \\
		R60 IMPLEMENT appropriate technical and organisational measures to protect personal data \\
		R63 PROTECT the personal data from unauthorised access and processing \\
		R65 IMPLEMENT a function to comply with local requirements and cross-border transfers \\
		\bottomrule
	\end{tabular}
\end{table}


\subsubsection{User participation}
All the requirements in this category specify the functionalities provided for data subjects to execute their individual rights in managing their personal data. The data subjects must be able to access and review, erase and rectify their personal data (e.g. R1, R44 and R45). The systems must allow the data subjects to withdraw consent (R6). The systems must also provide the data subjects their personal data when they would like to obtain and reuse their personal data for their own purposes across different services (R34).

\subsubsection{Notice}
It is the largest group consisting of 32 privacy requirements in the taxonomy. Most of the requirements in this category are concerned with the transparency of personal data processing (e.g. R17, R22 and R42). They aim to ensure that a system shall provide information related to the processing of personal data (e.g. what personal data is required, who is responsible for their data and results from requests) to data subjects.  Personal data shall not be misused, and the data subjects have the right to know the purpose of collection and processing (R38 and R39). The data subjects should be provided the duration their personal data will be stored (R55). Additional information must be provided to the data subjects if the collected personal data are required for other purposes (R37). General privacy-related information should be presented in a clear and simple, accessible language without technical terms as required in requirement R26. Any updates of personal data processing must be informed to the recipients (i.e. processors or third parties) of those personal data (R50). Some of the requirements in this category also belong to other categories such as the notices about data subjects' rights related to Category 1: user participation and the notices about data breaches related to Category 5: breach.

\subsubsection{User desirability}
The requirements in this category focus on ensuring that the processing of personal data is performed according to data subjects' consent and preferences. A number of requirements focus on the controllers being given authorities to process personal data (e.g. R8). It is also necessary to obtain consent for the processing based on those purposes (R35). The data subjects have options to allow the processing of their personal data for a certain specific purpose (R36).

According to GDPR Art. 6, in addition to consent, there are other ways for a lawful processing if it is necessary for the performance of a contract, compliance with a legal obligation, protect vital interests, the performance of task carried out in the public interest and the purposes of the legitimate interests. While R35 covers the consent aspects, we have also R39 in our taxonomy addresses the other aspects. R39 requires that the controllers, if not obtaining the consent, must provide data subjects the purpose(s) of the processing of personal data, including those listed in GDPR Art. 6.

\subsubsection{Data processing}
There are two requirements (R41 and R43) in this category, which change the traditional way of collecting and processing data. In the past, the data might be collected from data subjects as much as possible and kept in the system. They now require that the controllers are expected to collect and store only personal data that is required in the processing for the specific purpose(s). A set of requirements involves data erasure in systems. Requirement R53 addresses the case of removing personal data when the purpose for processing has expired. When the data subjects would like to have their personal data erased, the system shall provide this processing lawfully (R46 and R47). When the processing is complete and the personal data is no longer needed, the personal data should be removed from the system unless they are required by law/regulations (e.g. R51 and R52). In case that the personal data is unlawfully processed, the data must be removed from the systems (R7). In addition, personal data must be used only for the specified purpose(s) (R40). When requested by the data subjects, the controllers must transmit their personal data to another controller (R33). The data subject must be informed when their personal data needs to be transferred to a third country or an international organisation (R9). The controllers shall document the categories of personal data collected as it is important to know what personal data are stored in the systems (R70).

\subsubsection{Breach}
This goal category focuses on providing and notifying important information related to personal data breaches to data subjects, relevant stakeholders and a supervisory authority (e.g. R15 and R71). Thus, it is important to implement a functionality that satisfies this compliance in the systems (e.g. R66). Requirement R67 imposes good practices of informing the related parties about the breaches. The controllers must be informed by processors about the breaches as well (R68). The controllers shall document the details of data breaches for verifying their compliance (R69).

\subsubsection{Complaint/Request}
This privacy goal concerns complaint and request made by both data subjects and controllers. If the controllers refuse to take actions on the data subjects' requests about their individual rights, they have to provide a reason to the data subjects (R12). The data subjects must be able to lodge their complaints with a supervisory authority (R2). The controllers shall process personal data as requested (e.g. transmit personal data to another controller) (R33). The controllers should request additional information to confirm data subjects' identity when requests have been made (R31).

\subsubsection{Security}
There are thirteen requirements in our taxonomy covering the security practices in maintaining integrity, confidentiality and availability. The systems must allow only authorised people to access or process personal data (R56). The personal data should be protected with proper mechanisms (e.g. R60 and R63). The systems must restore the availability and access to personal data after incidents (R62). The interactions in the systems should neither identify nor observe the behaviour of the data subjects as well as reduce the linkability of the personal data collected (R64). Apart from the fundamental practices that the personal data should be protected, this is beneficial when the personal data is exposed. The data protection approaches such as anonymisation and pseudonymisation can help reduce the impact of privacy breaches. Most importantly, a set of requirements require the systems to implement mechanisms to ensure security and privacy compliance (e.g. R57, R58, R61 and R66). In addition, the implementation of mechanisms to assess the accuracy and quality of procedures should be considered (R49).

In case that the personal data are processed across organisations/countries, the controllers must ensure local requirements and cross-border transfers (R65). Cross-border transfers are challenging for both controllers and processors. In cross-border transfer settings, it is required that the requirements at the destination should be equivalent to the ones at the source. The processors outside EU are sometimes not aware of those scenarios since they may not normally process personal data of EU citizens and residents. Therefore, the controllers are responsible for verifying requirements compliance before transferring personal data.

%% file: sections/4-mining-new.tex
\section{Mining privacy requirements in issue reports} \label{sec4}

Most of today's software projects follow an agile style in which the software development is driven by resolving issues in the backlog. In those projects, issue reports contain information about requirements of a software that are recorded in multiple forms in an issue tracking system \cite{Choetkiertikul}: new requirements (such as user stories or new feature request issues), modification of existing requirements (such as improvement request issues) or reporting requirements not being properly met (i.e. bug report issues). Large projects may have thousands of issues which provide fairly comprehensive source of requirements about the projects and the associated software systems. Thus, we have performed a study on the issue reports of software projects to understand how the selected projects address the privacy requirements in our taxonomy. As issue reports can be in multiple forms, we note that the taxonomy can be applied to software requirements and user stories. In this section, we first describe the process steps that we have followed to carry out our study, assess classification results, resolve disagreements, and discuss the outcomes.

\begin{figure}[ht]
	\centering
	\includegraphics[width=1.0\linewidth]{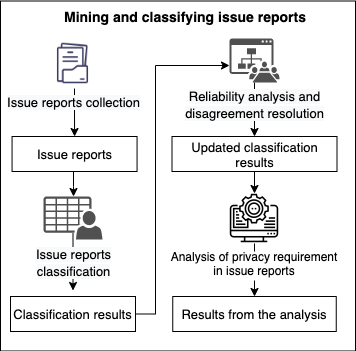}
	\caption{An overview process of mining and classifying privacy requirements in issue reports}
	\label{fig:Mining issue reports}
\end{figure}


\subsection{Issue reports collection}

We apply a number of criteria to select software projects for this study as follows: (i) are open-source projects, (ii) serve a large number of users, (iii) are related to privacy and (iv) have accessible issue tracking systems. There were a number of projects which satisfy these criteria. Among them, we selected Google Chrome and Moodle due to their large scale size, popularity and representativeness\footnote{As discussed later in the paper, performing thorough analysis as we have done on these two projects alone required significant effort (414 person-hours). Hence, we scoped our study in these two projects so that we can publish this dataset timely, enabling the community to initiate research on this important topic, and subsequently extend it with additional projects and data.}. Google Chrome is one of the most widely-used web browsers, which was developed under the Chromium Projects \cite{Projects}. As a web browser, Google Chrome stores personal data of users, e.g. username, password, address, credit card information, searching behaviour, history of visited sites and user location. Moodle is a well-known open-source learning platform \cite{Moodle} with over 100 million users worldwide. Moodle aims to comply with GDPR \cite{Moodle2019}.

Figure \ref{fig:Mining issue reports} shows the process steps that we have followed in our study. We first identified privacy-related issues from all the issue reports we collected from Google Chrome and Moodle projects. To do so, we identified issues that were explicitly tagged as privacy in the ``component'' field of both projects and their status was either assigned, fixed and verified (to ensure that they are valid issues). This is to ensure that the issue reports we collected were explicitly tagged as ``privacy'' by Chrome and Moodle contributors. This process initially gave us 1,080 privacy-related issues from Google Chrome and 524 from Moodle. We then manually examined those issues and filtered out those that have limited information (e.g. the description that does not explain the issue in detail or contain only source code) to enable us to perform the classification. For example, issue ID 953622\footnote{https://bugs.chromium.org/p/chromium/issues/detail?id=953622} states that ``Null-dereference READ in bool base::ContainsKey<std::\_\_Cr::map<std::\_\_Cr::basic\_string<char, std::\_\_Cr::c''. This issue description is very brief and does not contain any explanation describing what the issue is about, what personal data is concerned or which function is affected in the issue. We thus exclude the issue from our study. Finally, our data contains 896 issues from Chrome and 478 issues from Moodle.

In Chrome dataset, the collected issue reports were created between January 2009 and March 2020. There are five issue types reported in our dataset: bug, bug-regression, bug-security, feature and task. Each issue report has seven contributors on average; the contributors include reporters, owners and relevant collaborators. Issue reports in Moodle dataset span for two years, 2018 - 2019. The issue types in Moodle include bug, epic, improvement, new feature, task, sub-task and functional test. On average, five participants are involved in each report including reporters, assignees, testers and commenters. The descriptive statistics of the issue reports can be seen in Table \ref{tab:issue-stats}.

\begin{table*}[h]
	\centering
	\caption{Descriptive statistics of the number of contributors, resolution time and number of comments of the issue reports in our datasets.}
	\label{tab:issue-stats}
	\resizebox{6.5in}{!}{%
		\begin{tabular}{@{}llllllllllllllll@{}}
			\toprule
			\multirow{2}{*}{Project} & \multicolumn{5}{c}{\#Contributors} & \multicolumn{5}{c}{Resolution Time (days)} & \multicolumn{5}{c}{\#Comments} \\ \cmidrule(l){2-16}
			& min & max & mean & median & mode & min & max   & mean & median & mode & min & max & mean & median & mode \\ \midrule
			Google Chrome & 1   & 32  & 5    & 4      & 2    & 1   & 3,635 & 315  & 65     & 1    & 0   & 311 & 16   & 12     & 12    \\
			Moodle        & 1   & 14  & 4    & 5      & 5    & 1   & 852   & 37   & 13     & 1    & 0   & 112 & 11   & 9     & 1    \\ \bottomrule
		\end{tabular}%
	}
	{\parbox{16.5cm}{\footnotesize \#Contributors: number of contributors, \#Comments: number of comments, min: minimum of contributors/resolution time/comments, max: maximum of contributors/resolution time/comments, mean: mean of contributors/resolution time/comments, median: median of contributors/resolution time/comments, mode: mode of contributors/resolution time/comments.}}
\end{table*}

\subsection{Issue reports classification}

In this phase, we went through each issue report in the dataset to classify it into the privacy requirements in our taxonomy. This phase consists of three steps: (i) identifying concerned personal data described in the issue report, (ii) identifying functions/properties reported in the issue, (iii) mapping the issue to one or more privacy requirements. 

Regarding the classification, each coder was initially provided with an online form containing the title and description of the assigned issue reports and 71 columns representing each requirement. The coders analysed each issue following the classification steps described above (i.e. steps (i) - (iii)). The coders carefully consider every scenario mentioned in the issue reports. Once the coders have identified related information about personal data and function(s) concerned, they considered the relevant requirements. The coders determined the requirement(s) that matches with information analysed above. The coders then updated the value in the columns of chosen requirement(s) in the given form. Finally, all the coders delivered their result file containing the issue reports and their privacy requirement labels for reliability assessment process. To ensure that the classification process is reliable, two coders were assigned to classify an issue report. The reliability assessment and disagreement resolution processes are described in detail in Section \ref{4c}.

The following example demonstrates the issue reports classification in our datasets. Issue 123403\footnote{https://bugs.chromium.org/p/chromium/issues/detail?id=123403} in Google Chrome reports that \textit{``Regression: Can't delete individual cookies''}. The personal data affected here is individual cookies, and the function reported is erasing or deleting (individual cookies). Thus, we classify this issue into the requirement \textit{\textbf{ALLOW} the data subjects to erase their personal data (R44)} in our taxonomy (see Table \ref{tab:table1}). The users should be able to select the cookies that they want to delete.

In another example, issue 495226\footnote{https://bugs.chromium.org/p/chromium/issues/detail?id=495226} in Google Chrome requests that the \textit{``Change Sign-in confirmation screen''} should be changed. The description of this issue requires that the system should inform the reasons for user account data collection and how this data will be further processed before obtaining this data in the sign-in process. Since this issue requires that the user should be informed of the purpose of collection and processing, the issue can be classified into requirements R38 and R39 (see Table \ref{tab:table1}). Both of these requirements belong to the notice privacy goal. This example shows that an issue can be classified into more than one privacy requirements.

Issue 831572\footnote{https://bugs.chromium.org/p/chromium/issues/detail?id=831572} in Google Chrome requires: \textit{``Provide adequate disclosure for (potentially intrusive) policy configuration''}. Further investigation into the issue's description revealed that the disclosure of policy configuration includes: letting the users know that they are managed, and providing indication when user data may be intercepted and when user actions are logged locally. These involve the following functions: (i) the users should be informed of the purpose of processing so that they know they are managed; (ii) the enterprise may intercept the users' data, thus the users should know whom their personal data might be sent to; and (iii) the history of user logging shall be recorded to acquire logging data. Hence, this can be classified into three requirements R39, R27 and R13 (see Table \ref{tab:table1}). This example demonstrates that one issue relates to several requirements across different privacy goal categories: notice and data processing.

The following example demonstrates the issue that concerns multiple functions which more than one privacy goal is addressed in Moodle. Issue MDL-62904\footnote{https://tracker.moodle.org/browse/MDL-62904} in Moodle reports that \textit{``users can't find where to request account deletion''}. The issue was described that the system does not provide a function for users to request for deleting their account in the user interface. Hence, this issue addresses requirements R30 in the notice category and R44 in the user participation category in the taxonomy.

Although the mapping is relatively straightforward in most of the issues, some presents challenges. For instance, a set of issues in Moodle refer to the implementation of the core\_privacy plugins (e.g. MDL-61877\footnote{https://tracker.moodle.org/browse/MDL-61877}). However, the information given in the description of those issues is inadequate to identify which privacy requirements are related to. Therefore, we needed to seek for additional information about core\_privacy plugins in Moodle development documentation \cite{Moodle2019}. In addition, the issues in Google Chrome and Moodle projects have specific function names or technical terms used by their developers. Hence, extra effort was required to understand those issues and classify them. Once the coders acquired the additional information from software documentation, they discussed potential privacy issues/functionalities raised in those issue reports. For example, the documentation of core\_privacy plugins explains six functionalities that the plugins should provide. The coders then discussed which privacy requirements in the taxonomy involved with those functionalities, and classified that issue according to the identified privacy requirements.

\subsection{Reliability analysis} \label{4c}

The classification process has been performed by three coders, who also involved in the taxonomy development process. All the coders have independently followed the above process to classify issue reports into our taxonomy of privacy requirements. The first coder was responsible for classifying all 1,374 privacy issue reports. The classification process is however labour intensive. It took the first author approximately 138 person-hours to classify all 1,374 issue reports (6 minutes per issue report in average). In total, the process took four months to complete, so the coders had time to manage their workloads efficiently. The coders usually divided the assigned issue reports into smaller sets (i.e. about 50 issue reports) to work on for each round. For the purpose of reliability assessment, the second and third coder each was assigned to classify a different half of the issues in each project. This setting aimed to ensure that each issue report is classified by at least two coders.

An issue report can be classified into multiple privacy requirements (i.e. a multi-labelling problem). Hence, we employ Krippendorff's alpha coefficient \cite{Krippendorff2011}, \cite{Artstein2008} with MASI (Measurement Agreement on Set-valued Items) distance to measure agreement between coders with multi-label annotations \cite{Ravenscroft2016}, \cite{Passonneau2006}. The MASI distance measures difference between the sets of labels (i.e. privacy requirements) provided by two coders for a given issue. The Krippendorff's alpha values between the first and second coders are 0.509 for Chrome and 0.448 for Moodle. The agreement values between the first and third coder are 0.482 and 0.468 for Chrome and Moodle respectively. A disagreement resolution step was conducted to resolve the classification deviations between the three coders\footnote{Krippendorff’s alpha values indicate the degree of (dis-)agreement between coders. In our case, disagreements were often due to the ambiguity in the issue reports, requiring us to perform a resolution step. Doing these is to increase the reliability of our dataset.}.

\textbf{Disagreement resolution}: The low Krippendorff's alpha values from the initial classification were \emph{not} due to the privacy requirements in our taxonomy. Rather, it was mainly because of the limited information provided in the description of a number of issue reports, forcing the coders to make their own assumptions about the nature of those issues. If an issue report is clearly described, the coders classified it into the same requirements. 53.01\% of the Chrome issues and 46.23\% in Moodle received this total agreement between all the three coders. We have addressed this problem by conducting a disagreement resolution step where all the coders met and discussed to resolve the disagreements. 


We conducted several meeting sessions between the coders to resolve disagreements. Several meetings were conducted because of the time difference and availability among the coders. Each meeting took 3 hours on average as we revisited, discussed and reclassified the issues with disagreements issue by issue. During these meetings, the coders examined the issues thoroughly (not just only their description, but also other documentation related to the issues), discussed to develop a mutual understanding of the issue, and then reclassified the issue together. For each issue in the list, each coder explained the justification for their classification of the issue. The coders then resolved the disagreements on that issue in two ways. After the discussion, if the coders agreed with the other coder's classification, the labels were combined (i.e. the issue were classified into multiple requirements). If the coders did not reach an agreement, they went through the issue's description together to discuss and develop a mutual understanding of the issue. They then reclassified the issue together. Hence, the final classification has maximum agreement among the coders, thus ensuring the reliability of our dataset. Once all disagreements have been resolved for the sample set, the first coder adjusted the classification of the issues which are not included in the sample set to finalise the dataset. 

The following example illustrates how conflicts between coders were resolved. Issue 527346\footnote{https://bugs.chromium.org/p/chromium/issues/detail?id=527346} in Google Chrome requests that the users should know when they are managed. The description of the issue requires the system to show information to users when they are managed and the information should be easily seen by users. This issue was classified to R26 by one coder and R30 by the other coder. Although both R26 and R30 involve providing information to users, R26 focuses on the information relating to the policies, procedures, practices and logic of the processing of personal data, while R30 focuses on representing information relating to the processing of personal data with standardised icons in the user interface. After the coders revisited the issue and discussed the description in detail, the coders agreed that the information should be shown in the tray bubble which is a part of the user interface. The coders therefore reclassified this issue into R30.

Our work helps establish the traceability between issue reports and privacy concerns expressed in the regulations, standards and frameworks. These suggest what should be done to resolve the concerns raised in issue reports and meet the associated privacy requirements. In addition, this traceability facilitates compliance checking with respect to some specific regulations, standards and/or frameworks. The following examples demonstrate the privacy requirements traceability in issue reports. Referring to issue 123403 in Section 4.2, this bug-regression issue report concerns requirement R44 in the taxonomy. Requirement R44 was derived from GDPR, ISO/IEC 29100, Thailand PDPA and APEC privacy framework. To resolve this issue and comply with all four regulations, standards and frameworks, the system must provide a functionality for the controller to allow the data subjects to erase their personal data. In another example, referring to issue 123403 in Section 4.2, it addresses two privacy requirements, R30 and R44. Requirement R30 was derived from GDPR and APEC privacy framework. To resolve this issue and to comply with GDPR and APEC privacy framework, the system must also provide a functionality for the controller to provide the data subjects the information relating to the processing of personal data with standardised icons.

Our approach can be also applied to establish privacy requirements traceability in other software artifacts such as design models, source code and test cases. In general, our approach can be used a reference framework in developing privacy-aware software systems. For example, when a software engineer develops a system functionality which collects personal data, s/he may consult the taxonomy to identify the requirements related to the information required to provide to the data subjects before collecting their personal data such as concerned personal data (R42), purposes of collection (R38), purposes of processing (R39) and period/criteria used to store personal data (R55). The privacy requirements were derived from the regulations, standards and frameworks, hence meeting those requirements forms a basis for privacy compliance.

Although issue reports are a good source of information for software requirements, we acknowledge that they are not the only source. Software requirements can be in other forms such as requirements specifications or other documents (such as Confluence\footnote{https://www.atlassian.com/software/confluence/features} pages). The privacy requirements in the taxonomy can also be mapped to a list of privacy-related items concerned in a range of checks in organisations such as Data Protection Impact Assessment (DPIA). For example, DPIA requires a system to obtain consent from data subjects before processing their personal data. Requirement R35 in the taxonomy addresses this concern. However, the scope of this work focuses on privacy requirements in issue reports, and we plan to explore privacy requirements in other sources in our future work.

%% file: sections/5-analysis.tex
\section{Analysis and discussions} \label{results}

The study presented in the previous section generates not only a valuable dataset but also important insights into how privacy requirements have been addressed in Chrome and Moodle issue reports. In this section, we discuss and analyse some of the key findings and implications.


\subsection{The top and least concerned privacy requirements}

Table \ref{tab:table2} presents the coverage of each privacy goal in Chrome and Moodle issue reports. In both projects, the majority of the issues address the user participation requirements (Category 1). Most issue reports in Moodle address more than one privacy requirement across different privacy goal categories. This results in Moodle having higher coverage (in terms of the percentage of occurrences) than Chrome.

\begin{table}[ht]
	\centering
	\caption{The number of privacy requirements in each privacy goal and the percentage of occurrences in Google Chrome and Moodle issue reports}
	\label{tab:table2}
	\resizebox{3.25in}{!}{%
		\begin{tabular}{p{3cm} c c c}
			\toprule
			\multicolumn{1}{c}{\multirow{2}{*}{\textbf{Privacy goals}}} & \multicolumn{1}{c}{\multirow{2}{*}{\begin{tabular}[c]{@{}c@{}}\textbf{No. of privacy }\\\textbf{requirements~ }\end{tabular}}} & \multicolumn{2}{c}{\begin{tabular}[c]{@{}c@{}}\textbf{Percentage of}\\\textbf{occurrences (\%) }\end{tabular}} \\
			\cline{3-4}
			\multicolumn{1}{c}{} & \multicolumn{1}{c}{} & \multicolumn{1}{c}{\begin{tabular}[c]{@{}c@{}}\textbf{Google }\\\textbf{Chrome}\end{tabular}} & \multicolumn{1}{c}{\textbf{Moodle}} \\
			\midrule
			User participation & 9 & 35.83 & 68.62\\
			Notice & 32 & 30.36 & 47.91\\
			User desirability & 10 & 28.01 & 40.59\\
			Data processing & 16 & 13.06 & 11.51\\
			Breach & 6 & 0.00 & 0.00\\
			Complaint/Request & 5 & 0.11 & 1.88\\
			Security & 13 & 17.86 & 44.77\\
			\hline
			\multicolumn{4}{p{8.5cm}}{\footnotesize *Since one issue can relate to multiple privacy requirements, the sum of percentage exceeds 100\%} \setlength\lineskip{0pt}
		\end{tabular}%
	}
\end{table}

\begin{figure}
	\centering
	\includegraphics[width=1\linewidth]{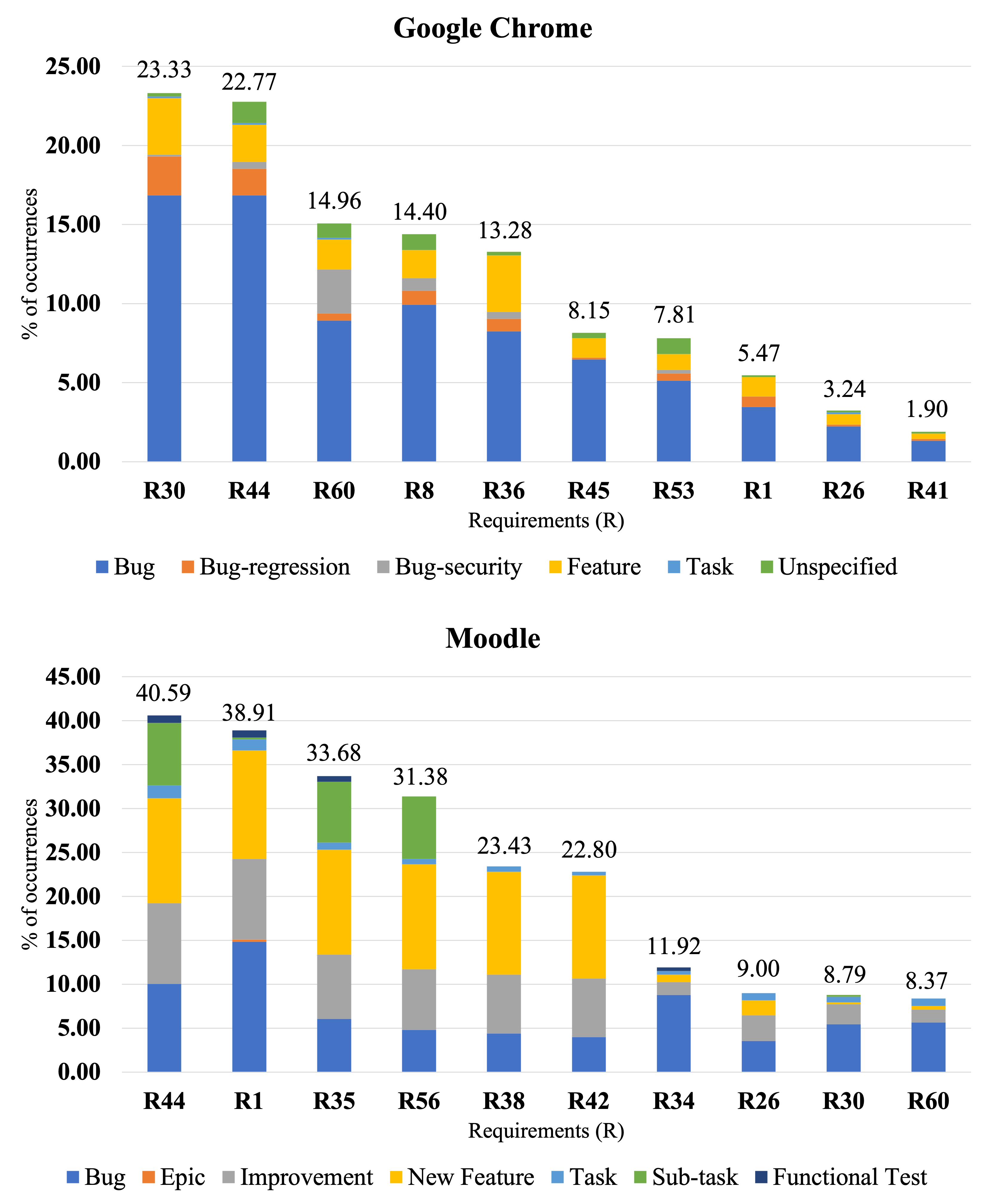}
	\caption{Top 10 privacy requirements occurrences in Google Chrome and Moodle datasets categorised by issue types}
	\label{fig:top10}
\end{figure}

The top three most concerned requirements in Chrome are R30, R44, R60 (see Figure \ref{fig:top10}). Note that these requirements belong to three different privacy goal categories (refer to Table \ref{tab:table1} for details of the privacy goals and requirements we discussed here). The top three requirements covered in Moodle issues are R44, R1 and R35 (see Figure \ref{fig:top10}). It is also worth noting that requirements R1, R26, R30, R44 and R60 are in the top 10 in both projects.

Requirement R44 was in the top three most concerned requirements in both projects, suggesting that allowing the data subjects to erase their personal data is a highly important privacy requirement for both Chrome and Moodle. Requirements R30 and R36 were also addressed in many privacy-related issues in Chrome. This suggests that providing information with standardised, visible and meaningful icons which inform the intended processing of personal data for users is an important privacy concern in Chrome (R30). In addition, many issues in Chrome also focus on addressing the privacy requirement that users are presented with all available options related to the processing of personal data (R36).

Apart from requirement R44, the other two requirements most frequently covered in Moodle issue reports are R1 and R35 (note that they are different from those in Chrome). Approximately 39\% issues in Moodle are related to requirement R1. This implies that Moodle has a strong emphasis on allowing users to access their personal data such as grade records, course participation and course enrolment records. This function is not only important for Moodle, but also in Chrome (R1 is also in the top 10 for Chrome). User consent is a major concern in privacy protection. We found that a large number of issue reports in Moodle explicitly requires the system to obtain consent from users for processing personal data based on specific purposes (R35).


It is interesting to note that none of the requirements in the breach category were satisfied by both Chrome and Moodle as they were not directly observed from issue reports or recorded in the ITS. These goals can be evidenced through high-level organisational activities such as Data Protection Impact Assessment, Legitimate Interest Assessment and breach notifications. Our future work will investigate this further.

We have performed further analysis on the issue reports that were classified into the top concerned requirements in Chrome and Moodle datasets. We analyse four factors focusing on how the contributors treat those issue reports: issue types, the time took to resolve an issue, the number of contributors involved and the number of comments associated with the issue.

In Chrome dataset, issue reports have five different types: bug, bug-regression, bug-security, feature and task. Bug type reports malfunctioning functionalities in current version of the system. Bug-regression focuses on the functions that used to work correctly in the previous versions, but are broken in the current version. Bug-security reports malfunctions that are risky to user security. Feature type requests for an implementation of a new function/feature. Task type, which is not a bug or feature, defines a piece of work that needs to be completed for an issue. A small group of issues were not specified their issue types. From our study, we found that the issue reports whose issue type is a bug are the largest group in every top concerned privacy requirement. These bug issue reports as well as bug-regression and bug-security took less time to resolve comparing to feature request issue reports on average for all the top concerned privacy requirements. In addition, the bug issue reports usually involved with a smaller number of contributors and had less discussions than the feature request issue reports. We have also investigated the discussions of bug issue reports classified into the top concerned requirements. We found that the bug issues were fixed after fifteen comments. However, the discussions of feature issues contain more details (e.g. use case scenarios, discussion points, screenshots and code snippets) than the bug issues.

There are seven different issue types in Moodle dataset: bug, epic, improvement, new feature, task, sub-task and functional test. The definitions of bug, new feature, task and sub-task issue types are similar to those mentioned in Chrome dataset. Epic issue type collects a group of issues that needs to be completed over a period of time. An improvement issue type is an enhancement to an existing feature. Functional test type contains the information and steps used for testing a particular function. From the analysis in Moodle dataset, the bug issue type contains the largest number of issue reports, followed by the feature issue type. We found that the bug issues did not only report the malfunctions, but they also reported the missing functionalities in the system (e.g. implement core\_privacy for block rss client). It is interesting to note that the new feature issues took less time to resolve comparing to the bug issues on average for all the requirements except R42. However, both issue types have similar number of comments (i.e. 11 to 12 comments) and number of contributors (i.e. 5).

\input{sections/RQ-privacy-vs-nonprivacy}

%% file: sections/RQ-privacy-vs-nonprivacy.tex

\subsection{The treatment of privacy and non-privacy issues}

We investigate if privacy issues were treated differently from non-privacy issues in Chrome and Moodle. We focus on observing two kinds of treatments: the time it took to resolve an issue and the number of comments associated with the issue. The former reflects how fast an issue was resolved while the latter indicates the attention and engagement of the project team to the issue. We randomly sampled the dataset we built earlier using a 95\% confidence level with a confidence interval of 5\footnote{https://www.surveysystem.com/sscalc.htm} to obtain 269 privacy issues from Chrome and 213 from Moodle. Applying the same sampling scheme, we randomly selected 382 non-privacy issues from Chrome and 380 from Moodle - these issues were not tagged as privacy in the ``component'' field. Note that the resolution time is calculated from the number of days between reported date and the date when the issue was flagged as being resolved.

\begin{table}
	\centering
	\caption{Results of the Wilcoxon rank-sum test: non-privacy vs. privacy issues}
	\label{tab:ranksum}
	\resizebox{8.5cm}{!}{
		\begin{tabular}{l l l l l}
			\toprule
			\textbf{Project} & \textbf{Attribute} & \textbf{One-sided tail} & \textbf{p-value} & \textbf{Effect size}\\
			\midrule
			Google Chrome & Resolution time & Less & $<$0.001 & 0.578 \\
			Google Chrome & \#Comments & Less & $<$0.001 & 0.691 \\
			Moodle & Resolution time & Greater & $<$0.001 & 0.609 \\
			Moodle & \#Comments & Greater & $<$0.001 & 0.604\\
			\bottomrule
		\end{tabular}%
	}	
\end{table}

We employ the Wilcoxon rank-sum test (also known as Mann-Whitney U test), a non-parametric hypothesis test which compares the difference between two independent observations \cite{Wild1997}. We performed two tests between privacy and non-privacy samples, one for the resolution time and the other for the number of comments. The results (see Table \ref{tab:ranksum}) show that the resolution time and the number of comments are statistically significantly (\textit{p-value $\leq$ 0.001}) different between privacy and non-privacy issues in both Chrome and Moodle with effect size greater than 0.5 in all cases.

We also compare the median rank of the two samples using one-tailed test. Our results show that privacy issues were resolved more quickly and attracted less comments than non-privacy issues in Moodle (see Table \ref{tab:ranksum}). On the other hand, it took longer to resolve privacy issues than non-privacy issues in Chrome. Also, privacy issues in Chrome tend to attract more discussion than non-privacy issues. 

We observed that the contributors in Chrome were confident and had more experience in resolving non-privacy issues. Hence, these issues attracted less discussion and was resolved more quickly. By contrast, privacy issues in Chrome attracted more discussions since the contributors were uncertain about the issues and their affected components in the system. We observed five examples that the contributors commented in those privacy issues as follows: (i) the contributors did not know what the affected components reported in the issues do (e.g. issue 345741); (ii) the contributors could not identify the causes of issues; (iii) the contributors required time and effort to come up with potential solutions; (iv) the contributors needed to assess the difficulties of the issues and their resolutions; and (v) the contributors did not know whom to assign the work. These reasons also led to longer time to resolve the privacy issues in Chrome. In addition, the Chrome project does not have a well-defined process that specifically handles privacy. Hence, the contributors need to ensure that fixing privacy issues will not create another problem in different components. Thus, resolving privacy issues attracted a lot of discussions, leading to longer resolution time.

On the other hand, privacy issues in Moodle were resolved more quickly and attracted less comments than non-privacy issues. We observe that the privacy issues were well reported and clearly explained in Moodle. Moodle contributors were familiar with privacy-related functionalities and relevant system components. In addition, Moodle has a clearly defined infrastructure to handle privacy and privacy compliance in the system (e.g. privacy API \cite{Moodle2019} and GDPR for plugin developers \cite{Nicols2018}). This infrastructure includes a number of components that support privacy-related functionalities and several key individual rights in GDPR (e.g. accessing to personal data and requesting for deletion). When there is a privacy-related bug or new feature request, the contributors can consult the privacy API documentation and identify the components that they must fix or implement. Hence, the privacy issues took less time and attracted less comments in Moodle.

%% file: sections/6-discussion.tex
\section{Threats to validity} \label{threats}


Our study involved subjective judgements. We have applied several strategies to mitigate this threat such as using multiple coders (who are the authors of the paper), applying inter-rater reliability assessments, organising training sessions and disagreement resolution meetings. A legal expert could have extended the view of the human coders in interpreting legal regulations. However, we note that all the coders had attended a training course on privacy regulations (including GDPR). This has enhanced our interpretation of privacy regulations from a legal perspective, thus minimised this risk. In addition, our study was built upon well-founded processes and theories in previous work such as Grounded Theory \cite{Glaser2017} and Goal-Based Requirements Analysis Method (GBRAM) \cite{Antn2004},\cite{Anton1996}. We also used relevant statistical measures and techniques to ensure that our findings are statistically significant.


Our taxonomy of requirements is derived from the GDPR and ISO/IEC 29100 -- two major and widely-adopted privacy and data protection regulations and framework to date; and Thailand PDPA and APEC privacy framework -- two region-specific representative regulations and privacy frameworks. We are aware that there are other privacy protection laws and regulations. However, most of them share many commonalities with the GDPR and ISO/IEC 29100 as confirmed by Thailand PDPA and APEC privacy framework in this study. In fact, previous studies (e.g. \cite{Ayala-Rivera2018}, \cite{Linden2020}, \cite{Tsohou2020}) have shown that GDPR is one of the most comprehensive data protection regulations. GDPR provides a more overarching coverage compared to other data protection regulations in terms of privacy concerns (i.e. principles and rights) and organisational impacts. Many other privacy laws have been developed as inspired by GDPR (e.g. Brazil's LGPD and CCPA in California) \cite{GDPR.EU2020}. It was also used as a benchmark for other countries to develop data protection regulations \cite{Torre} such as Japan, South Korea, Thailand and other countries in Asia Pacific (APAC -- Asia Pacific Privacy Authorities) \cite{Laboris2019}. Hence, although the principles or rights in other laws and regulations are slightly different due to variations in each country/city, they share many commonalities with GDPR (e.g. see the comparison between GDPR and CCPA in \cite{Iubenda2020},\cite{DataPrivacyManager}). ISO/IEC 29100 has also been used to develop organisational and technical privacy controls in many information and communication systems \cite{PECB2015}. Therefore, we found that GDPR and  ISO/IEC 29100 together are the most comprehensive, thus our taxonomy can generalise to other privacy regulations and standards. We acknowledge that future research could involve investigating country specific privacy regulations, and extending our taxonomy accordingly. We also note that developing a taxonomy of privacy requirements for GDPR, ISO/IEC 29100, Thailand PDPA and APEC privacy framework already required substantial efforts and detailed processes. To the best our knowledge, our work is the first comprehensive taxonomy of privacy requirements analysing both data protection regulations and privacy frameworks. This would lay foundations for future research in this timely, important area of software engineering.

Finally, we performed the mining and classification of issue reports for Google Chrome and Moodle. These are two large and widely-used software systems that have strong emphasis on privacy concerns. However, we acknowledge that our datasets may not be representative of other software applications. Further investigation is required to explore other projects in different domains (e.g. e-health software systems and mobile applications). However, we note that building this dataset on two projects alone required substantial effort and highly thorough processes (414 person-hours).

%% file: sections/7-conclusions.tex
\section{Conclusion and future work} \label{sec7}

Privacy compliance in software systems has become tremendously significant for organisations that handle personal data of their users. In this paper, we have developed a comprehensive taxonomy of privacy requirements based on four well-known standardised data protection regulations and privacy frameworks (GDPR, ISO/IEC 29100, Thailand PDPA and APEC privacy framework). Our approach is built upon a content analysis process which is adapted from the Goal-Based Requirements Analysis Method and based on Grounded Theory \cite{Antn2004}, \cite{Anton1996}. Thus, the steps in this approach (privacy requirements identification, privacy requirements refinement and privacy requirements classification) are generic and applicable to different regulations and privacy standards. In our approach, privacy requirements are formed by three components: actions, affected parties/objects and target results. These are commonly found in the narrative statements across different regulations. 

We also performed reliability assessments and disagreement resolution in the process to ensure that our taxonomy is reliably constructed. Our taxonomy consists of 71 privacy requirements grouped in 7 privacy goal categories. Since the studied regulations and frameworks are not specific to any software types, our taxonomy is generally applicable to a wide range of software applications. In this study, we retain the privacy requirements at this level since the implementation can vary depending on organisational structures, software architectures, existing implementation and knowledge, and the aptitude of software development teams. Thus, our taxonomy focuses on defining privacy requirements a software needs to satisfy, leaving how the software is designed and implemented to meet those requirements to project stakeholders.

We have also performed a study on how two large projects (Google Chrome and Moodle) address those privacy requirements in our taxonomy. To do so, we mined the issue reports recorded in those projects and collected 1,374 privacy-related issues. We then classified these issues into our taxonomy through a process which involved multiple coders and the use of inter-rater reliability assessments and disagreement resolution. We found that the privacy requirements in the user participation category were covered in a majority of the issues, while none of the issues were found to address the breach category. The taxonomy also captures all the privacy requirements in the issue reports. We also found that the time taken to resolve privacy-related issues and the degree of developers' engagement on them were also statistically significantly different from those of non-privacy issues. We note that the taxonomy can be applied to software requirements and user stories. In fact, there are issue reports in the dataset which are user stories or requirements for new features.

We believe that the taxonomy and the dataset are important contributions to the community (given that none exists). This would initiate discussions and further work (e.g. refining and/or extending the taxonomy, performing user studies, etc.) in the community. Our work lays several important foundations for future research in this area. The systematic method performed in the work enables future research to be conducted on other data protection and privacy regulations and frameworks. The taxonomy can act as a reference for the research community to discuss and expand. We plan to investigate other privacy regulations and policies and extend our taxonomy, if necessary. The requirements in our taxonomy are written in natural language and structured into templates. Although we believe that this is the most intuitive form to developers, future work could explore other alternative forms such as semantic frame-based representation \cite{Bhatia2019}. We have manually derived requirements in this study as it is essential to examine structure of statements and how privacy requirements are expressed in different regulations and frameworks. However, the requirements extraction process can be automated using natural language processing (NLP) techniques to identify actors, actions, objects and other relevant constraints from text documents (i.e. regulations and frameworks). These methods have been used in several studies (e.g. \cite{Mu2009} and \cite{Li}). Legal experts could also be involved to help interpret legal perspective of the taxonomy in the future work.

Future work also involves exploring how issue reports in other projects (e.g. health and mobile applications) attend to the requirements in our taxonomy. In addition, software developers would be able to participate in validating the use of taxonomy and their understanding towards the derived privacy requirements. Furthermore, we plan to develop tool support to automate this classification task. Other potential future work includes the investigation of the use of the taxonomy with semantic web technologies to facilitate computational and regulatory risk analysis purposes. The investigation of other forms of traceability relationships such as the traceability between code and issue reports or code and privacy requirements can also be further studied. 

%% file: TSE22-privacy-main-majRev-wh.bbl
\begin{thebibliography}{94}
\providecommand{\natexlab}[1]{#1}
\providecommand{\url}[1]{#1}
\csname url@samestyle\endcsname
\providecommand{\newblock}{\relax}
\providecommand{\bibinfo}[2]{#2}
\providecommand{\BIBentrySTDinterwordspacing}{\spaceskip=0pt\relax}
\providecommand{\BIBentryALTinterwordstretchfactor}{4}
\providecommand{\BIBentryALTinterwordspacing}{\spaceskip=\fontdimen2\font plus
\BIBentryALTinterwordstretchfactor\fontdimen3\font minus
  \fontdimen4\font\relax}
\providecommand{\BIBforeignlanguage}[2]{{%
\expandafter\ifx\csname l@#1\endcsname\relax
\typeout{** WARNING: IEEEtranN.bst: No hyphenation pattern has been}%
\typeout{** loaded for the language `#1'. Using the pattern for}%
\typeout{** the default language instead.}%
\else
\language=\csname l@#1\endcsname
\fi
#2}}
\providecommand{\BIBdecl}{\relax}
\BIBdecl

\bibitem[{Office Journal of the European
  Union}(2016)]{OfficeJournaloftheEuropeanUnion;2016}
\BIBentryALTinterwordspacing
{Office Journal of the European Union}, ``{Regulation (EU) 2016/679 of the
  European Parliament and of the Council of 27 April 2016 on the protection of
  natural persons with regard to the processing of personal data and on the
  free movement of such data, and repealing Directive 95/46/EC (General Da},''
  Tech. Rep., 2016. [Online]. Available:
  \url{https://eur-lex.europa.eu/legal-content/EN/TXT/PDF/?uri=OJ:L:2016:119:FULL}
\BIBentrySTDinterwordspacing

\bibitem[ISO/IEC(2011)]{ISO/IEC2011}
ISO/IEC, ``{International Standard ISO/IEC 29100},'' ISO/IEC, Tech. Rep., 2011.

\bibitem[Ayala-Rivera and Pasquale(2018)]{Ayala-Rivera2018}
V.~Ayala-Rivera and L.~Pasquale, ``{The grace period has ended: An approach to
  operationalize GDPR requirements},'' in \emph{Proceedings - 2018 IEEE 26th
  International Requirements Engineering Conference, RE 2018}.\hskip 1em plus
  0.5em minus 0.4em\relax Institute of Electrical and Electronics Engineers
  Inc., oct 2018, pp. 136--146.

\bibitem[Solove(2006)]{Solove}
\BIBentryALTinterwordspacing
D.~J. Solove, ``{A taxonomy of privacy},'' \emph{University of Pennsylvania Law
  Review}, vol. 154, no.~3, pp. 477--564, 2006. [Online]. Available:
  \url{https://scholarship.law.gwu.edu/faculty{\_}publications}
\BIBentrySTDinterwordspacing

\bibitem[Kalloniatis et~al.(2008)Kalloniatis, Kavakli, and
  Gritzalis]{Kalloniatis2008}
C.~Kalloniatis, E.~Kavakli, and S.~Gritzalis, ``{Addressing privacy
  requirements in system design: The PriS method},'' \emph{Requirements
  Engineering}, vol.~13, no.~3, pp. 241--255, aug 2008.

\bibitem[Deng et~al.(2011)Deng, Wuyts, Scandariato, Preneel, and
  Joosen]{Deng2011}
\BIBentryALTinterwordspacing
M.~Deng, K.~Wuyts, R.~Scandariato, B.~Preneel, and W.~Joosen, ``{A privacy
  threat analysis framework: Supporting the elicitation and fulfillment of
  privacy requirements},'' \emph{Requirements Engineering}, vol.~16, no.~1, pp.
  3--32, mar 2011. [Online]. Available:
  \url{https://link.springer.com/article/10.1007/s00766-010-0115-7}
\BIBentrySTDinterwordspacing

\bibitem[Heurix et~al.(2015)Heurix, Zimmermann, Neubauer, and Fenz]{Heurix2015}
\BIBentryALTinterwordspacing
J.~Heurix, P.~Zimmermann, T.~Neubauer, and S.~Fenz, ``{A taxonomy for privacy
  enhancing technologies},'' \emph{Computers and Security}, vol.~53, pp. 1--17,
  jun 2015. [Online]. Available:
  \url{http://dx.doi.org/10.1016/j.cose.2015.05.002}
\BIBentrySTDinterwordspacing

\bibitem[Perera et~al.(2016)Perera, McCormick, Bandara, Price, and
  Nuseibeh]{Perera2016}
C.~Perera, C.~McCormick, A.~K. Bandara, B.~A. Price, and B.~Nuseibeh,
  ``{Privacy-by-design framework for assessing internet of things applications
  and platforms},'' in \emph{ACM International Conference Proceeding Series},
  2016, pp. 83--92.

\bibitem[Aljeraisy et~al.(2020)Aljeraisy, Rana, and Perera]{Aljeraisy2020}
\BIBentryALTinterwordspacing
A.~Aljeraisy, O.~Rana, and C.~Perera, ``{A Systematic Analysis of Privacy Laws
  and Privacy by Design Schemes for the Internet of Things: A Developer's
  Perspective},'' aug 2020. [Online]. Available:
  \url{https://hal.archives-ouvertes.fr/hal-02567959v3
  https://hal.archives-ouvertes.fr/hal-02567959/document}
\BIBentrySTDinterwordspacing

\bibitem[G{\"{u}}rses et~al.(2011)G{\"{u}}rses, Troncoso, and Diaz]{Gurses2011}
S.~G{\"{u}}rses, C.~Troncoso, and C.~Diaz, ``{Engineering Privacy by Design},''
  2011.

\bibitem[Senarath and Arachchilage(2018{\natexlab{a}})]{Senarath2018b}
A.~Senarath and N.~A. Arachchilage, ``{Why developers cannot embed privacy into
  software systems?}'' 2018.

\bibitem[Bednar et~al.(2019)Bednar, Spiekermann, and Langheinrich]{Bednar2019}
\BIBentryALTinterwordspacing
K.~Bednar, S.~Spiekermann, and M.~Langheinrich, ``{Engineering Privacy by
  Design: Are engineers ready to live up to the challenge?}'' \emph{Information
  Society}, vol.~35, no.~3, pp. 122--142, 2019. [Online]. Available:
  \url{https://www.tandfonline.com/action/journalInformation?journalCode=utis20}
\BIBentrySTDinterwordspacing

\bibitem[Cavoukian(2009)]{Cavoukian2009}
A.~Cavoukian, ``{Privacy by Design - The 7 foundational principles -
  Implementation and mapping of fair information practices},''
  \emph{Information and Privacy Commissioner of Ontario, Canada}, p.~10, 2009.

\bibitem[van Rest et~al.(2014)van Rest, Boonstra, Everts, van Rijn, and van
  Paassen]{Rest2014}
J.~van Rest, D.~Boonstra, M.~Everts, M.~van Rijn, and R.~van Paassen,
  ``{Designing privacy-by-design},'' in \emph{Lecture Notes in Computer Science
  (including subseries Lecture Notes in Artificial Intelligence and Lecture
  Notes in Bioinformatics)}, vol. 8319.\hskip 1em plus 0.5em minus 0.4em\relax
  Springer Verlag, oct 2014, pp. 55--72.

\bibitem[G{\"{u}}rses and {Del {\'{A}}lamo}(2016)]{Gurses2016}
S.~G{\"{u}}rses and J.~M. {Del {\'{A}}lamo}, ``{Privacy Engineering: Shaping an
  Emerging Field of Research and Practice},'' \emph{IEEE Security and Privacy},
  vol.~14, no.~2, pp. 40--46, mar 2016.

\bibitem[Spiekermann(2012)]{Spiekermann2012}
S.~Spiekermann, ``{The challenges of privacy by design},'' pp. 38--40, jul
  2012.

\bibitem[Mikkelsen et~al.(2019)Mikkelsen, Soller, Strandell-Jansson, and
  Wahlers]{Mikkelsen2019}
\BIBentryALTinterwordspacing
D.~Mikkelsen, H.~Soller, M.~Strandell-Jansson, and M.~Wahlers, ``{GDPR
  compliance challenges since May 2018},'' Tech. Rep., 2019. [Online].
  Available:
  \url{https://www.mckinsey.com/business-functions/risk/our-insights/gdpr-compliance-after-may-2018-a-continuing-challenge{\#}}
\BIBentrySTDinterwordspacing

\bibitem[{Capgemini Research Institute}(2019)]{Capgemini2019}
\BIBentryALTinterwordspacing
{Capgemini Research Institute}, ``{Championing Data Protection and Privacy: a
  source of competitive advantage in the digital century.}'' Tech. Rep., 2019.
  [Online]. Available:
  \url{https://www.capgemini.com/wp-content/uploads/2019/09/Report{\_}Championing-Data-Protection-and-Privacy.pdf}
\BIBentrySTDinterwordspacing

\bibitem[Hadar et~al.(2018)Hadar, Hasson, Ayalon, Toch, Birnhack, Sherman, and
  Balissa]{Hadar2018}
I.~Hadar, T.~Hasson, O.~Ayalon, E.~Toch, M.~Birnhack, S.~Sherman, and
  A.~Balissa, ``{Privacy by designers: software developers' privacy mindset},''
  \emph{Empirical Software Engineering}, vol.~23, no.~1, pp. 259--289, 2018.

\bibitem[Swinhoe(2020)]{Swinhoe2020}
\BIBentryALTinterwordspacing
D.~Swinhoe, ``{The 14 biggest data breaches of the 21st century},'' 2020.
  [Online]. Available:
  \url{https://www.csoonline.com/article/2130877/the-biggest-data-breaches-of-the-21st-century.html}
\BIBentrySTDinterwordspacing

\bibitem[Ant{\'{o}}n et~al.(2002)Ant{\'{o}}n, Earp, and Reese]{Anton2002}
A.~I. Ant{\'{o}}n, J.~B. Earp, and A.~Reese, ``{Analyzing Website privacy
  requirements using a privacy goal taxonomy},'' in \emph{Proceedings of the
  IEEE International Conference on Requirements Engineering}, 2002.

\bibitem[Ant{\'{o}}n and Earp(2004)]{Antn2004}
A.~I. Ant{\'{o}}n and J.~B. Earp, ``{A requirements taxonomy for reducing Web
  site privacy vulnerabilities},'' \emph{Requirements Engineering}, vol.~9,
  no.~3, pp. 169--185, 2004.

\bibitem[Anthonysamy et~al.(2017)Anthonysamy, Rashid, and
  Chitchyan]{Anthonysamy2017}
P.~Anthonysamy, A.~Rashid, and R.~Chitchyan, ``{Privacy requirements: Present
  {\&} future},'' in \emph{Proceedings - 2017 IEEE/ACM 39th International
  Conference on Software Engineering: Software Engineering in Society Track,
  ICSE-SEIS 2017}.\hskip 1em plus 0.5em minus 0.4em\relax Institute of
  Electrical and Electronics Engineers Inc., jun 2017, pp. 13--22.

\bibitem[Guarda and Zannone(2009)]{Guarda2009}
P.~Guarda and N.~Zannone, ``{Towards the development of privacy-aware
  systems},'' \emph{Information and Software Technology}, vol.~51, no.~2, pp.
  337--350, feb 2009.

\bibitem[Choetkiertikul et~al.(2021)Choetkiertikul, Dam, Tran, Pham,
  Ragkhitwetsagul, and Ghose]{Choetkiertikul}
\BIBentryALTinterwordspacing
M.~Choetkiertikul, H.~K. Dam, T.~Tran, T.~Pham, C.~Ragkhitwetsagul, and
  A.~Ghose, ``{Automatically recommending components for issue reports using
  deep learning},'' \emph{Empirical Software Engineering}, vol.~26, no.~2,
  2021. [Online]. Available: \url{https://doi.org/10.1007/s10664-020-09898-5}
\BIBentrySTDinterwordspacing

\bibitem[Moodle(2021)]{MoodleTracker}
\BIBentryALTinterwordspacing
Moodle, ``{Tracker introduction},'' 2021. [Online]. Available:
  \url{https://docs.moodle.org/dev/Tracker{\_}introduction}
\BIBentrySTDinterwordspacing

\bibitem[Moodle(2017)]{MoodleFeature}
\BIBentryALTinterwordspacing
------, ``{New feature ideas - MoodleDocs},'' 2017. [Online]. Available:
  \url{https://docs.moodle.org/dev/New{\_}feature{\_}ideas}
\BIBentrySTDinterwordspacing

\bibitem[Merten et~al.(2015)Merten, Mager, H{\"{u}}ubner, Quirchmayr, Paech,
  and B{\"{u}}rsner]{Merten2015}
T.~Merten, B.~Mager, P.~H{\"{u}}ubner, T.~Quirchmayr, B.~Paech, and
  S.~B{\"{u}}rsner, ``{Requirements communication in issue tracking systems in
  four open-source projects},'' \emph{CEUR Workshop Proceedings}, vol. 1342,
  pp. 114--125, 2015.

\bibitem[Merten et~al.(2016)Merten, Falis, H{\"{u}}bner, Quirchmayr,
  B{\"{u}}rsner, and Paech]{Merten2016}
T.~Merten, M.~Falis, P.~H{\"{u}}bner, T.~Quirchmayr, S.~B{\"{u}}rsner, and
  B.~Paech, ``{Software Feature Request Detection in Issue Tracking Systems},''
  in \emph{Proceedings - 2016 IEEE 24th International Requirements Engineering
  Conference, RE 2016}.\hskip 1em plus 0.5em minus 0.4em\relax Institute of
  Electrical and Electronics Engineers Inc., dec 2016, pp. 166--175.

\bibitem[Ant{\'{o}}n(1996)]{Anton1996}
A.~I. Ant{\'{o}}n, ``{Goal-based requirements analysis},'' in \emph{Proceedings
  of the IEEE International Conference on Requirements Engineering}.\hskip 1em
  plus 0.5em minus 0.4em\relax IEEE, 1996, pp. 136--144.

\bibitem[Sangaroonsilp et~al.(2021)Sangaroonsilp, Dam, Choetkiertikul,
  Ragkhitwetsagul, and Ghose]{reppkg-pridp}
\BIBentryALTinterwordspacing
P.~Sangaroonsilp, H.~K. Dam, M.~Choetkiertikul, C.~Ragkhitwetsagul, and
  A.~Ghose, ``{Replication Package for Mining and Classifying Privacy and Data
  Protection Requirements in Issue Reports manuscript},'' 2021. [Online].
  Available: \url{https://bit.ly/Mining-privacy-reqs-rev22}
\BIBentrySTDinterwordspacing

\bibitem[Beckers(2012)]{Beckers2012}
K.~Beckers, ``{Comparing privacy requirements engineering approaches},'' in
  \emph{Proceedings - 2012 7th International Conference on Availability,
  Reliability and Security, ARES 2012}, 2012, pp. 574--581.

\bibitem[Danezis(2007)]{George2007}
G.~Danezis, ``{Introduction to Privacy Technology},'' Tech. Rep., 2007.

\bibitem[Solove(2008)]{Solove2008}
\BIBentryALTinterwordspacing
D.~J. Solove, ``{Understanding Privacy},'' 2008. [Online]. Available:
  \url{https://papers.ssrn.com/abstract=1127888}
\BIBentrySTDinterwordspacing

\bibitem[Pfitzmann and Hansen(2010)]{anon_terminology}
\BIBentryALTinterwordspacing
A.~Pfitzmann and M.~Hansen, ``{A terminology for talking about privacy by data
  minimization: Anonymity, Unlinkability, Undetectability, Unobservability,
  Pseudonymity, and Identity Management},'' apr 2010. [Online]. Available:
  \url{http://dud.inf.tu-dresden.de/literatur/Anon{\_}Terminology{\_}v0.33.pdf}
\BIBentrySTDinterwordspacing

\bibitem[{British Standards Institute}(2000)]{BritishStandardsInstitute2000}
{British Standards Institute}, ``{ISO 17799 (2000) Information technology code
  of practice for information security management},'' Tech. Rep., 2000.

\bibitem[Spiekermann and Cranor(2009)]{Spiekermann2009}
S.~Spiekermann and L.~F. Cranor, ``{Engineering privacy},'' \emph{IEEE
  Transactions on Software Engineering}, vol.~35, no.~1, pp. 67--82, 2009.

\bibitem[OECD(2013)]{OECD2013}
OECD, ``{The OECD Privacy Framework},'' Tech. Rep., 2013.

\bibitem[Meis et~al.(2015)Meis, Wirtz, and Heisel]{Meis}
R.~Meis, R.~Wirtz, and M.~Heisel, ``{A taxonomy of requirements for the privacy
  goal transparency},'' in \emph{Lecture Notes in Computer Science (including
  subseries Lecture Notes in Artificial Intelligence and Lecture Notes in
  Bioinformatics)}, vol. 9264, 2015, pp. 195--209.

\bibitem[Meis and Heisel(2016)]{Meis2016}
R.~Meis and M.~Heisel, ``{Understanding the privacy goal intervenability},'' in
  \emph{Lecture Notes in Computer Science (including subseries Lecture Notes in
  Artificial Intelligence and Lecture Notes in Bioinformatics)}, vol. 9830
  LNCS, 2016, pp. 79--94.

\bibitem[Breaux(2014)]{Breaux2014}
T.~Breaux, ``{Privacy requirements in an age of increased sharing},''
  \emph{IEEE Software}, vol.~31, no.~5, pp. 24--27, 2014.

\bibitem[Senarath and Arachchilage(2018{\natexlab{b}})]{Senarath2018}
A.~R. Senarath and N.~A.~G. Arachchilage, ``{Understanding user privacy
  expectations: A software developer's perspective},'' \emph{Telematics and
  Informatics}, 2018.

\bibitem[Sheth et~al.(2014)Sheth, Kaiser, and Maalej]{Sheth2014}
\BIBentryALTinterwordspacing
S.~Sheth, G.~Kaiser, and W.~Maalej, ``{Us and them: A study of privacy
  requirements across north america, asia, and Europe},'' in \emph{Proceedings
  - International Conference on Software Engineering}, no.~1, 2014, pp.
  859--870. [Online]. Available:
  \url{http://mobis.informatik.uni-hamburg.de/privacy-}
\BIBentrySTDinterwordspacing

\bibitem[Birnhack et~al.(2014)Birnhack, Toch, and Hadar]{Birnhack2014}
M.~Birnhack, E.~Toch, and I.~Hadar, ``{Privacy Mindset, Technological
  Mindset},'' \emph{SSRN Electronic Journal}, vol.~55, no. Fall, pp. 1--46,
  2014.

\bibitem[Linden et~al.(2020)Linden, Khandelwal, Harkous, and Fawaz]{Linden2020}
\BIBentryALTinterwordspacing
T.~Linden, R.~Khandelwal, H.~Harkous, and K.~Fawaz, ``{The privacy policy
  landscape after the GDPR},'' pp. 47--64, sep 2020. [Online]. Available:
  \url{https://content.sciendo.com/view/journals/popets/2020/1/article-p47.xml}
\BIBentrySTDinterwordspacing

\bibitem[Wilson et~al.(2016)Wilson, Schaub, Dara, Liu, Cherivirala, Leon,
  Andersen, Zimmeck, Sathyendra, Russell, Norton, Hovy, Reidenberg, and
  Sadeh]{Wilson2016}
S.~Wilson, F.~Schaub, A.~A. Dara, F.~Liu, S.~Cherivirala, P.~G. Leon, M.~S.
  Andersen, S.~Zimmeck, K.~M. Sathyendra, N.~C. Russell, T.~B. Norton, E.~Hovy,
  J.~Reidenberg, and N.~Sadeh, ``{The creation and analysis of a Website
  privacy policy corpus},'' in \emph{54th Annual Meeting of the Association for
  Computational Linguistics, ACL 2016 - Long Papers}, vol.~3, 2016, pp.
  1330--1340.

\bibitem[Gruschka et~al.(2018)Gruschka, Mavroeidis, Vishi, and
  Jensen]{Gruschka2018}
N.~Gruschka, V.~Mavroeidis, K.~Vishi, and M.~Jensen, ``{Privacy issues and data
  protection in big data: A case study analysis under GDPR},'' in \emph{2018
  IEEE International Conference on Big Data (Big Data)}.\hskip 1em plus 0.5em
  minus 0.4em\relax Seattle, WA, USA: IEEE, 2018, pp. 5027--5033.

\bibitem[M{\"{u}}ller et~al.(2019)M{\"{u}}ller, Kowatsch, Debus, Mirdita, and
  B{\"{o}}ttinger]{Muller2019}
\BIBentryALTinterwordspacing
N.~M. M{\"{u}}ller, D.~Kowatsch, P.~Debus, D.~Mirdita, and K.~B{\"{o}}ttinger,
  ``{On GDPR Compliance of Companies' Privacy Policies},'' in \emph{Lecture
  Notes in Computer Science (including subseries Lecture Notes in Artificial
  Intelligence and Lecture Notes in Bioinformatics)}.\hskip 1em plus 0.5em
  minus 0.4em\relax Springer Verlag, sep 2019, vol. 11697, pp. 151--159.
  [Online]. Available: \url{https://doi.org/10.1007/978-3-030-27947-9{\_}13}
\BIBentrySTDinterwordspacing

\bibitem[Torre et~al.(2020)Torre, Abualhaija, Sabetzadeh, Briand, Baetens,
  Goes, and Forastier]{Torre}
D.~Torre, S.~Abualhaija, M.~Sabetzadeh, L.~Briand, K.~Baetens, P.~Goes, and
  S.~Forastier, ``{An AI-assisted Approach for Checking the Completeness of
  Privacy Policies Against GDPR},'' \emph{2020 IEEE 28th International
  Requirements Engineering Conference (RE)}, pp. 136--146, 2020.

\bibitem[Torre et~al.(2019)Torre, Soltana, Sabetzadeh, Briand, Auffinger, and
  Goes]{Torre2019}
D.~Torre, G.~Soltana, M.~Sabetzadeh, L.~C. Briand, Y.~Auffinger, and P.~Goes,
  ``{Using Models to Enable Compliance Checking against the GDPR: An Experience
  Report},'' in \emph{Proceedings - 2019 ACM/IEEE 22nd International Conference
  on Model Driven Engineering Languages and Systems, MODELS 2019}, 2019, pp.
  1--11.

\bibitem[Breaux et~al.(2014)Breaux, Hibshi, and Rao]{Breaux2014a}
T.~D. Breaux, H.~Hibshi, and A.~Rao, ``{Eddy, a formal language for specifying
  and analyzing data flow specifications for conflicting privacy
  requirements},'' \emph{Requirements Engineering}, vol.~19, no.~3, pp.
  281--307, 2014.

\bibitem[Bhatia et~al.(2019)Bhatia, Evans, and Breaux]{Bhatia2019}
\BIBentryALTinterwordspacing
J.~Bhatia, M.~C. Evans, and T.~D. Breaux, ``{Identifying incompleteness in
  privacy policy goals using semantic frames},'' \emph{Requirements
  Engineering}, vol.~24, no.~3, pp. 291--313, 2019. [Online]. Available:
  \url{https://doi.org/10.1007/s00766-019-00315-y}
\BIBentrySTDinterwordspacing

\bibitem[Pandit et~al.(2019)Pandit, Polleres, Bos, Brennan, Bruegger, Ekaputra,
  Fern{\'{a}}ndez, Hamed, Kiesling, Lizar, Schlehahn, Steyskal, and
  Wenning]{Pandit2019}
\BIBentryALTinterwordspacing
H.~J. Pandit, A.~Polleres, B.~Bos, R.~Brennan, B.~Bruegger, F.~J. Ekaputra,
  J.~D. Fern{\'{a}}ndez, R.~G. Hamed, E.~Kiesling, M.~Lizar, E.~Schlehahn,
  S.~Steyskal, and R.~Wenning, ``{Creating a vocabulary for data privacy: The
  first-year report of data privacy vocabularies and controls community group
  (DPVCG)},'' in \emph{Lecture Notes in Computer Science (including subseries
  Lecture Notes in Artificial Intelligence and Lecture Notes in
  Bioinformatics)}, vol. 11877 LNCS.\hskip 1em plus 0.5em minus 0.4em\relax
  Springer, Cham, oct 2019, pp. 714--730. [Online]. Available:
  \url{https://link.springer.com/chapter/10.1007/978-3-030-33246-4{\_}44}
\BIBentrySTDinterwordspacing

\bibitem[de~Carvalho et~al.(2020)de~Carvalho, {Del Prete}, Martin, {Araujo
  Rivero}, {\"{O}}nen, Schiavo, Rum{\'{i}}n, Mouratidis, Yelmo, and
  Koukovini]{EUcluster2020}
\BIBentryALTinterwordspacing
R.~M. de~Carvalho, C.~{Del Prete}, Y.~S. Martin, R.~M. {Araujo Rivero},
  M.~{\"{O}}nen, F.~P. Schiavo, {\'{A}}.~C. Rum{\'{i}}n, H.~Mouratidis, J.~C.
  Yelmo, and M.~N. Koukovini, ``{Protecting Citizens' Personal Data and
  Privacy: Joint Effort from GDPR EU Cluster Research Projects},'' \emph{SN
  Computer Science}, vol.~1, no.~4, p. 217, 2020. [Online]. Available:
  \url{https://doi.org/10.1007/s42979-020-00218-8}
\BIBentrySTDinterwordspacing

\bibitem[BPR4GDPR()]{BPR4GDPR}
\BIBentryALTinterwordspacing
BPR4GDPR, ``{Business Process Re-engineering and functional toolkit for GDPR
  compliance}.'' [Online]. Available: \url{https://www.bpr4gdpr.eu/}
\BIBentrySTDinterwordspacing

\bibitem[DEFeND()]{DEFEND}
\BIBentryALTinterwordspacing
DEFeND, ``{What is the Defend Project - Defend Project}.'' [Online]. Available:
  \url{https://www.defendproject.eu/}
\BIBentrySTDinterwordspacing

\bibitem[SMOOTH()]{SMOOTH}
\BIBentryALTinterwordspacing
SMOOTH, ``{SMOOTH platform - GDPR Compliance Cloud Platform for Micro
  Enterprises}.'' [Online]. Available: \url{https://smoothplatform.eu/}
\BIBentrySTDinterwordspacing

\bibitem[PDP4E()]{PDP4E}
\BIBentryALTinterwordspacing
PDP4E, ``{PDP4E Project | European Project}.'' [Online]. Available:
  \url{https://www.pdp4e-project.eu/}
\BIBentrySTDinterwordspacing

\bibitem[Ferreyra et~al.(2020)Ferreyra, Tessier, Pedroza, and
  Heisel]{Ferreyra2020}
\BIBentryALTinterwordspacing
N.~E. Ferreyra, P.~Tessier, G.~Pedroza, and M.~Heisel, ``{PDP-ReqLite: A
  Lightweight Approach for the Elicitation of Privacy and Data Protection
  Requirements},'' in \emph{Lecture Notes in Computer Science (including
  subseries Lecture Notes in Artificial Intelligence and Lecture Notes in
  Bioinformatics)}, vol. 12484 LNCS.\hskip 1em plus 0.5em minus 0.4em\relax
  Springer Science and Business Media Deutschland GmbH, sep 2020, pp. 161--177.
  [Online]. Available: \url{https://doi.org/10.1007/978-3-030-66172-4{\_}10}
\BIBentrySTDinterwordspacing

\bibitem[Information(2017)]{POSEIDON}
\BIBentryALTinterwordspacing
S.~Information, ``{Protection and control of Secured Information by means of a
  privacy enhanced Dashboard},'' pp. 1--75, 2017. [Online]. Available:
  \url{https://www.poseidon-h2020.eu/}
\BIBentrySTDinterwordspacing

\bibitem[{United Nations Conference on Trade and Development
  (UNCTAD)}(2020)]{UNCTAD2020}
\BIBentryALTinterwordspacing
{United Nations Conference on Trade and Development (UNCTAD)}, ``{Data
  Protection and Privacy Legislation Worldwide},'' 2020. [Online]. Available:
  \url{https://unctad.org/page/data-protection-and-privacy-legislation-worldwide}
\BIBentrySTDinterwordspacing

\bibitem[Li et~al.(2019)Li, Yu, and He]{Li1}
H.~Li, L.~Yu, and W.~He, ``{The Impact of GDPR on Global Technology
  Development},'' \emph{Journal of global information technology management :
  JGITM}, vol.~22, no.~1, pp. 1--6, 2019.

\bibitem[{Data Privacy Manager}()]{Data}
\BIBentryALTinterwordspacing
{Data Privacy Manager}, ``{5 biggest GDPR fines so far [2020] – Data Privacy
  Manager}.'' [Online]. Available:
  \url{https://dataprivacymanager.net/5-biggest-gdpr-fines-so-far-2020/}
\BIBentrySTDinterwordspacing

\bibitem[{European Commission}(2019)]{EuropeanCommission2019}
\BIBentryALTinterwordspacing
{European Commission}, ``{GDPR IN NUMBERS 2019 {\#}HAPPYBIRTHDAYGDPR},'' Tech.
  Rep., 2019. [Online]. Available:
  \url{https://ec.europa.eu/info/sites/info/files/infographic-gdpr{\_}in{\_}numbers.pdf}
\BIBentrySTDinterwordspacing

\bibitem[CNET(2019)]{CNET}
\BIBentryALTinterwordspacing
CNET, ``{British Airways faces {\$}230M GDPR fine for 2018 data breach},''
  2019. [Online]. Available: \url{https://cnet.co/2EHboiu}
\BIBentrySTDinterwordspacing

\bibitem[{Privacy Affairs}(2020)]{PrivacyAffa}
\BIBentryALTinterwordspacing
{Privacy Affairs}, ``{GDPR Fines List: Find all GDPR fines {\&} detailed
  statistics},'' 2020. [Online]. Available:
  \url{https://www.privacyaffairs.com/gdpr-fines/}
\BIBentrySTDinterwordspacing

\bibitem[{National Legislative Assembly}(2019)]{PDPA}
\BIBentryALTinterwordspacing
{National Legislative Assembly}, ``{The Thailand Personal Data Protection
  Act},'' Tech. Rep., 2019. [Online]. Available:
  \url{http://www.ratchakitcha.soc.go.th/DATA/PDF/2562/A/069/T{\_}0052.PDF}
\BIBentrySTDinterwordspacing

\bibitem[{Asia-Pacific Economic Cooperation (APEC)}(2017)]{Apec2015}
{Asia-Pacific Economic Cooperation (APEC)}, ``{APEC Privacy Framework
  (2015)},'' APEC Secretariat, Tech. Rep., 2017.

\bibitem[Glaser and Strauss(2017)]{Glaser2017}
B.~Glaser and A.~L. Strauss, \emph{{The discovery of grounded theory :
  strategies for qualitative research}}.\hskip 1em plus 0.5em minus 0.4em\relax
  Routledge, 2017.

\bibitem[Anton and Potts(1998)]{Anton1998}
A.~I. Anton and C.~Potts, ``{Use of goals to surface requirements for evolving
  systems},'' in \emph{Proceedings - International Conference on Software
  Engineering}.\hskip 1em plus 0.5em minus 0.4em\relax IEEE Comp Soc, 1998, pp.
  157--166.

\bibitem[Baumer et~al.(2000)Baumer, Earp, and Payton]{Baumer2000}
D.~Baumer, J.~B. Earp, and F.~C. Payton, ``{Privacy of Medical Records : IT
  Implications of HIPAA},'' \emph{ACM SIGCAS Computers and Society}, vol.~30,
  no. December, pp. 40--47, dec 2000.

\bibitem[Viera and Garrett(2005)]{Viera2005}
A.~J. Viera and J.~M. Garrett, ``{Understanding interobserver agreement: the
  kappa statistic},'' \emph{Family Medicine}, vol.~37, no.~5, pp. 360--363,
  2005.

\bibitem[Hallgren(2012)]{Hallgren}
K.~A. Hallgren, ``{Computing inter-rater reliability for observational data: An
  overview and tutorial.}'' \emph{Tutorials in Quantitative Methods for
  Psychology}, vol.~8, no.~1, pp. 23--34, 2012.

\bibitem[Fleiss(1971)]{Fleiss1971}
J.~L. Fleiss, ``{Measuring nominal scale agreement among many raters},''
  \emph{Psychological Bulletin}, vol.~76, no.~5, pp. 378--382, nov 1971.

\bibitem[Landis and Koch(1977)]{Landis1977}
J.~R. Landis and G.~G. Koch, ``{The Measurement of Observer Agreement for
  Categorical Data},'' Tech. Rep.~1, 1977.

\bibitem[Kateifides et~al.()Kateifides, Potter, Highams, Young, Kazmi,
  Campbell, Ashcroft, Campbell, Kerpauskaite, Dampster, Filis, Richter,
  Formichella, Jamallsawat, Mcnair, and Brikshasri]{Kateifides}
A.~Kateifides, A.~Potter, H.~Highams, A.~Young, T.~Kazmi, C.~Campbell,
  V.~Ashcroft, K.~Campbell, K.~Kerpauskaite, E.~Dampster, A.~Filis, B.~Richter,
  J.~P. Formichella, N.~Jamallsawat, B.~Mcnair, and A.~Brikshasri, ``{Comparing
  privacy laws: GDPR v. Thai Personal Data Protection Act},'' Tech. Rep.

\bibitem[{State of California Department of
  Justice}(2018)]{StateofCaliforniaDepartmentofJustice2018}
\BIBentryALTinterwordspacing
{State of California Department of Justice}, ``{California Consumer Privacy Act
  (CCPA)},'' 2018. [Online]. Available: \url{https://oag.ca.gov/privacy/ccpa}
\BIBentrySTDinterwordspacing

\bibitem[{The Chromium Projects}()]{Projects}
\BIBentryALTinterwordspacing
{The Chromium Projects}, ``{The Chromium Projects}.'' [Online]. Available:
  \url{https://www.chromium.org/chromium-projects}
\BIBentrySTDinterwordspacing

\bibitem[Moodle()]{Moodle}
\BIBentryALTinterwordspacing
Moodle, ``{About Moodle - MoodleDocs}.'' [Online]. Available:
  \url{https://docs.moodle.org/38/en/About{\_}Moodle}
\BIBentrySTDinterwordspacing

\bibitem[Moodle(2019)]{Moodle2019}
\BIBentryALTinterwordspacing
------, ``{Privacy API},'' 2019. [Online]. Available:
  \url{https://docs.moodle.org/dev/Privacy{\_}API}
\BIBentrySTDinterwordspacing

\bibitem[Krippendorff(2011)]{Krippendorff2011}
\BIBentryALTinterwordspacing
K.~Krippendorff, ``{Computing Krippendorff's Alpha-Reliability},'' Tech. Rep.,
  2011. [Online]. Available: \url{http://repository.upenn.edu/asc{\_}papers/43}
\BIBentrySTDinterwordspacing

\bibitem[Artstein and Poesio(2008)]{Artstein2008}
\BIBentryALTinterwordspacing
R.~Artstein and M.~Poesio, ``{Inter-coder agreement for computational
  linguistics},'' pp. 555--596, 2008. [Online]. Available:
  \url{www.essex.ac.uk/Research/nle/arrau/}
\BIBentrySTDinterwordspacing

\bibitem[Ravenscroft et~al.(2016)Ravenscroft, Oellrich, Saha, and
  Liakata]{Ravenscroft2016}
\BIBentryALTinterwordspacing
J.~Ravenscroft, A.~Oellrich, S.~Saha, and M.~Liakata, ``{Multi-label annotation
  in scientific articles - The multi-label cancer risk assessment corpus},'' in
  \emph{Proceedings of the 10th International Conference on Language Resources
  and Evaluation, LREC 2016}, 2016, pp. 4115--4123. [Online]. Available:
  \url{https://www.aclweb.org/anthology/L16-1650}
\BIBentrySTDinterwordspacing

\bibitem[Passonneau(2006)]{Passonneau2006}
R.~Passonneau, ``{Measuring agreement on set-valued items (MASI) for semantic
  and pragmatic annotation},'' in \emph{Proceedings of the 5th International
  Conference on Language Resources and Evaluation, LREC 2006}, 2006, pp.
  831--836.

\bibitem[Wild(1997)]{Wild1997}
\BIBentryALTinterwordspacing
C.~Wild, ``{The Wilcoxon Rank-Sum Test},'' Department of Statistics, University
  of Auckland, Tech. Rep., 1997. [Online]. Available:
  \url{https://www.stat.auckland.ac.nz/{~}wild/ChanceEnc/Ch10.wilcoxon.pdf}
\BIBentrySTDinterwordspacing

\bibitem[Nicols(2018)]{Nicols2018}
A.~Nicols, ``{GDPR For Plugin Developers: Moodle's Privacy API},'' Tech. Rep.,
  2018.

\bibitem[Tsohou et~al.(2020)Tsohou, Magkos, Mouratidis, Chrysoloras, Piras,
  Pavlidis, Debussche, Rotoloni, and Crespo]{Tsohou2020}
A.~Tsohou, M.~Magkos, H.~Mouratidis, G.~Chrysoloras, L.~Piras, M.~Pavlidis,
  J.~Debussche, M.~Rotoloni, and B.~G.~N. Crespo, ``{Privacy, security, legal
  and technology acceptance requirements for a gdpr compliance platform},'' in
  \emph{Lecture Notes in Computer Science (including subseries Lecture Notes in
  Artificial Intelligence and Lecture Notes in Bioinformatics)}, vol. 11980
  LNCS, 2020, pp. 204--223.

\bibitem[GDPR.EU(2020)]{GDPR.EU2020}
\BIBentryALTinterwordspacing
GDPR.EU, ``{How the GDPR could change in 2020},'' 2020. [Online]. Available:
  \url{https://gdpr.eu/gdpr-in-2020/}
\BIBentrySTDinterwordspacing

\bibitem[Laboris(2019)]{Laboris2019}
\BIBentryALTinterwordspacing
I.~Laboris, ``{The impact of the GDPR outside the EU},'' 2019. [Online].
  Available:
  \url{https://www.lexology.com/library/detail.aspx?g=872b3db5-45d3-4ba3-bda4-3166a075d02f}
\BIBentrySTDinterwordspacing

\bibitem[Iubenda(2020)]{Iubenda2020}
\BIBentryALTinterwordspacing
Iubenda, ``{CCPA vs GDPR: what's the difference?}'' 2020. [Online]. Available:
  \url{https://www.iubenda.com/en/help/21109-ccpa-vs-gdpr}
\BIBentrySTDinterwordspacing

\bibitem[{Data Privacy Manager}(2020)]{DataPrivacyManager}
\BIBentryALTinterwordspacing
{Data Privacy Manager}, ``{CCPA vs. GDPR - differences and similarities},''
  2020. [Online]. Available: \url{https://dataprivacymanager.net/ccpa-vs-gdpr/}
\BIBentrySTDinterwordspacing

\bibitem[PECB(2015)]{PECB2015}
\BIBentryALTinterwordspacing
PECB, ``{ISO 29100 How Can Organizations Secure Its Privacy Network?}'' 2015.
  [Online]. Available:
  \url{https://pecb.com/whitepaper/iso-29100--how-can-organizations-secure-its-privacy-network}
\BIBentrySTDinterwordspacing

\bibitem[Mu et~al.(2009)Mu, Wang, and Guo]{Mu2009}
Y.~Mu, Y.~Wang, and J.~Guo, ``{Extracting software functional requirements from
  free text documents},'' in \emph{2009 International Conference on Information
  and Multimedia Technology, ICIMT 2009}, 2009, pp. 194--198.

\bibitem[Li et~al.(2015)Li, Guzman, Tsiamoura, Schneider, and Bruegge]{Li}
Y.~Li, E.~Guzman, K.~Tsiamoura, F.~Schneider, and B.~Bruegge, ``{Automated
  requirements extraction for scientific software},'' in \emph{Procedia
  Computer Science}, vol.~51, no.~1, 2015, pp. 582--591.

\end{thebibliography}
